%% file: main.tex
\documentclass[12pt]{article}

\usepackage{geometry}
\usepackage{amssymb,amsmath,amsfonts,eurosym,ulem,graphicx,caption,color,setspace,sectsty,comment,natbib,pdflscape, array}
\usepackage{hyperref}
\hypersetup{
colorlinks=true,
citecolor=blue,
linkcolor=black,
filecolor=black,      
urlcolor=black,
}
\usepackage{tikz-qtree}
\usepackage{bbold}
\usepackage[bottom]{footmisc}
\usepackage{algorithm, algpseudocode}
\usepackage{tabularx}
\usepackage{caption}
\usepackage{subcaption}
\usepackage{hyperref}
\usepackage{float}
\usepackage{graphicx}
\usepackage{booktabs}
\usepackage[flushleft]{threeparttable}
\usepackage{bm}

\newcommand{\bX}{\mathbf{X}}

\usepackage{scalerel,stackengine}
\usepackage{color}
\allowdisplaybreaks
\stackMath
\newcommand\reallywidehat[1]{%
	\savestack{\tmpbox}{\stretchto{%
			\scaleto{%
				\scalerel*[\widthof{\ensuremath{#1}}]{\kern-.6pt\bigwedge\kern-.6pt}%
				{\rule[-\textheight/2]{1ex}{\textheight}}
			}{\textheight}%
		}{0.5ex}}%
	\stackon[1pt]{#1}{\tmpbox}%
}

\usepackage{scalerel}
\usepackage{stackengine,wasysym}

\makeatletter
\newcommand*{\indep}{%
	\mathbin{%
		\mathpalette{\@indep}{}%
	}%
}
\newcommand*{\nindep}{%
	\mathbin{
		\mathpalette{\@indep}{\not}
	}%
}
\newcommand*{\@indep}[2]{%
	\sbox0{$#1\perp\m@th$}
	\sbox2{$#1=$}
	\sbox4{$#1\vcenter{}$}
	\rlap{\copy0}
	\dimen@=\dimexpr\ht2-\ht4-.2pt\relax
	\kern\dimen@
	{#2}%
	\kern\dimen@
	\copy0 
} 
\makeatother

\newcolumntype{L}[1]{>{\raggedright\let\newline\\arraybackslash\hspace{0pt}}m{#1}}
\newcolumntype{C}[1]{>{\centering\let\newline\\arraybackslash\hspace{0pt}}m{#1}}
\newcolumntype{R}[1]{>{\raggedleft\let\newline\\arraybackslash\hspace{0pt}}m{#1}}

\begin{document}

\begin{titlepage}
\title{\Large Assessing Sensitivity of Machine Learning Predictions.\\ A Novel Toolbox with an Application to Financial Literacy.\footnotetext{We are grateful to participants at KU Leuven and IMT School for Advanced Studies seminars and at the 2020 Data Science Webinar. We want to thank Tommaso Agasisti, Ilja Corneliz, Chiara Masci, Giorgio Gnecco, Chris Van Klaveren, Kwonsang Lee, Fabrizia Mealli, Rachel Nethery, Massimo Riccaboni, Armando Rungi, and Mike Smet. Falco J. Bargagli-Stoffi acknowledges funding from the Alfred P. Sloan Foundation. Kenneth De Beckker acknowledges the financial support from Wikifin.be. Joana Elisa Maldonado acknowledges support from the Flemish Science Organization through the grant S000617N. Kristof De Witte acknowledges the financial support from Wikifin.be and from the Flemish Science Organization through the grant S000617N.}}
\author{\normalsize Falco J. Bargagli-Stoffi\thanks{\scriptsize Corresponding author. Mail to: \href{mailto:fbargaglistoffi@hsph.harvard.edu}{fbargaglistoffi@hsph.harvard.edu}. Harvard University, 655 Huntington Ave, Boston, MA 02115, United States.} \and \normalsize Kenneth De Beckker\thanks{\scriptsize Hasselt University, Agoralaan Building D, 3590 Diepenbeek, Belgium. KU Leuven, Warmoesberg 26 - 1000 Brussels, Belgium. } \and \normalsize Joana Elisa Maldonado\thanks{\scriptsize KU Leuven, Naamsestraat 69 - 3000 Leuven, Belgium.} \and \normalsize Kristof De Witte\thanks{\scriptsize KU Leuven, Naamsestraat 69 - 3000 Leuven, Belgium. UNU-Merit, Maastricht University, Minderbroedersberg 4 - 6211 LK Maastricht, The Netherlands.} }
\date{}
\maketitle
\vspace{-1.25cm}
\begin{abstract}
\noindent{Despite their popularity, machine learning predictions are sensitive to potential unobserved predictors. This paper proposes a general algorithm that assesses how the omission of an unobserved variable with high explanatory power could affect the predictions of the model. Moreover, the algorithm extends the usage of machine learning from pointwise predictions to inference and sensitivity analysis. In the application, we show how the framework can be applied to data with inherent uncertainty, such as students' scores in a standardized assessment on financial literacy. First, using Bayesian Additive Regression Trees (BART), we predict students' financial literacy scores (FLS) for a subgroup of students with missing FLS. Then, we assess the sensitivity of predictions by comparing the predictions and performance of models with and without a highly explanatory synthetic predictor. We find no significant difference in the predictions and performances of the \textit{augmented} (i.e., the model with the synthetic predictor) and \textit{original} model. This evidence sheds a light on the stability of the predictive model used in the application. The proposed methodology can be used -- above and beyond our motivating empirical example -- in a wide range of machine learning applications in social and health sciences.}   \\ 

\vspace{0in}
\end{abstract}
\setcounter{page}{0}
\thispagestyle{empty}
\end{titlepage}

\doublespacing

\section{Introduction} \label{sec:introduction}

Machine learning is rapidly growing in popularity in social sciences as it helps scholars to address a large set of issues regarding prediction, causal inference, theory testing and development of new data sources \citep{mullainathan2017machine, athey2019machine}. This rise has been mainly driven by the staggering performance of machine learning in predictive tasks. However, the usage of off-the-shelf machine learning methodologies to perform predictions should be done in a careful and thoughtful way. To this day, little has been done to explore how sensitive machine learning predictions are to potentially unobserved predictors. 

From a methodological point of view, this paper accounts and accommodates for this shortcoming by proposing a general algorithm that assesses how the omission of an unobserved predictor could affect the predictions and the performance of the model. As such, we obtain estimates for, what we call, \textit{sensitivity of prediction analysis}.  The algorithm that we propose is general enough to  be applied to both Bayesian methodologies \cite[i.e., the Bayesian Additive Regression Tree method introduced by][]{chipman2010bart} and frequentist techniques \citep[i.e., the random forest methodology by][]{breiman2001random}. In particular, our method aims at assessing the extent to which an unobserved predictor would impact the model's predictions and its performance.

To perform this analysis, we generate a \textit{synthetic} predictor with a high explanatory power (i.e., high correlation with the outcome) but that is uncorrelated with all the predictors in the model. Then, we check how the inclusion of this additional variable in the machine learning model modifies its predictions and accuracy. We assume that the model at hand is able to catch most of the \textit{signal} from the data if its forecasts and its predictive ability are not extensively modified by the inclusion of the synthetic predictor. In order to assess this, we check if: (i) the unit level predictions of the model with the synthetic predictor -- what we call the \textit{augmented model} -- fall within the confidence interval of the unit level prediction of the model that we are comparing; and (ii) the predictive performance of the augmented model is not statistically different from the performance of the original model. The approach that we develop here is applied in the case of a continuous outcome variable, but it can be easily extended to a binary outcome variable.  Moreover, we make this methodology robust by introducing, in a second step, a set of correlations between the synthetic predictor and the most important predictors in the original model. This second case more closely mimics real-world applications where predictors are correlated. 

On the applied side, we innovate the literature by introducing uncertainty of predictions, which has various applications as machine learning is increasingly used to perform predictions on sensitive manners: e.g., patients’ life expectancy \citep{kleinberg2015prediction}, human capital selection \citep{chalfin2016productivity}, unemployment insurance effectiveness \citep{chernozhukov2017double}, students' learning abilities \citep{kotsiantis2004predicting, delen2010comparative, kotsiantis2012use}, and so on. In this spirit, the proposed methodology is a first step towards a more transparent way to account for predictions' stability and uncertainty in machine learning applications in social and health sciences.

As a motivating application, we are the first to introduce supervised machine learning in the financial literacy literature. First, we use machine learning to predict financial literacy scores (FLS) for students in a region of Belgium where these scores are not observed. We train our model using the PISA data \citep{OECD2017a} for the Flemish region of Belgium were the FLS are observed. Then, we perform the prediction for the missing FLS of students in the Walloon region of Belgium. We use an extensive set of student and school-level variables as predictors in our machine learning model. Second, we use machine learning to identify the characteristics of students with outlying predicted test scores. In our case, the main focus is on students with lower predicted FLS. Indeed, machine learning can help researchers and policy-makers to understand the characteristics of students who have low predictions for financial literacy. These insights could enable policy-makers to target interventions to lower performing students in order to improve their performances. Third, we employ the novel sensitivity analysis to assess the robustness of our analyses. Sensitivity is central in the case of our application, as we deal with sensitive data such as students' scores in a standardized assessment on financial literacy.  In this setting, it is deemed important to assess how sensitive the predictions of students' FLS are to a potentially unobserved predictor with high explanatory power.

Our main results can be summarized as follows. The performance of the proposed model is good as we reach an estimated adjusted $R^2$ of roughly 73\% and both the root mean squared error (RMSE) and mean absolute error (MAE) of prediction are comparatively small.\footnote{This performance measures result from 10-folds cross-validation. K-folds cross-validation is a well known way to assess the performance of a predictive model in machine learning \citep{james2013tree}.} Moreover, the best performing machine learning technique, Bayesian Additive Regression Trees \citep{chipman2010bart}, allows us to get draws from the posterior distribution of the predicted observations and construct credible intervals for the unit level predictions. Hence, we are able to detect the predictions that fall outside the 95\% credible intervals for the mean posterior predicted values. We use this information to identify a set of outlier predictions. On one side, this analysis shows that the predictive  probability  of  having  a  low  financial  literacy  score  is  the  largest for  students  with  lower  scores for  reading  and math  in  the  PISA  test. On the other side, the students' background plays a critical role: students with the largest predicted low FLS are often individuals from families where the school language is not spoken at home and the parents have a poor educational background.
Moreover, results from the sensitivity of predictions analysis show that the inclusion of an unobserved predictor would not affect the predictions and the predictive ability of the model in a sizable manner. This hints at the fact that our model is already getting enough \textit{signal} to perform its predictions.

The remainder of the paper is organized as follows. In Section \ref{sec:methodology} we introduce the machine learning methodologies used and developed in this paper. Section \ref{sec:data} describes the PISA data used to illustrate the proposed methodology. In Section \ref{sec:results} we discuss the results obtained from the application on the PISA data. Section \ref{sec:discussion} concludes with a discussion of the results.\footnote{The \texttt{R} and \texttt{Stata} codes used for the analysis, together with the data and the functions for the machine learning analysis are publicly available on the GitHub page of the corresponding author (\href{https://github.com/fbargaglistoffi/sensitivity-analysis-machine-learning}{\texttt{https://github.com/fbargaglistoffi/sensitivity-analysis-machine-learning}}).}

\section{Methodology}\label{sec:methodology}

This Section discusses in detail the machine learning technique used for prediction (Subsection \ref{subsec:bart}) and its extension to the detection of observations with outlying predicted values which paves the way to targeted policies. In Subsection \ref{subsec:sensitivity}, we introduce the novel methodologies for sensitivity analysis that extend the usage of machine learning algorithms from pointwise predictions to robustness analysis regarding these predictions. 

\subsection{Machine learning for predictions and targeted policies}\label{subsec:bart}

We rely on Bayesian Additive Regression Tree (BART) algorithm \citep{chipman2010bart} -- which is a Bayesian ensemble of trees methodology -- as a predictive model.  BART was shown to have an excellent performance both in prediction tasks \citep{murray2017log, linero2018bayesian1, linero2018bayesian2, hernandez2018bayesian, bargagli2020machine, bargagli2020supervised} and in causal inference tasks \citep{hill2011bayesian, logan2019decision, nethery2019estimating, bargagli2019heterogeneous, bargagli2020essays, lee2020causal}. This is due to the particular flexibility of this method and the fact that it overcomes potential issues connected with other machine learning methodologies such as the Classification And Regression Tree (CART) algorithm \citep{friedman1984classification} and the Random Forest (RF) algorithm \citep{breiman2001random}. Moreover, BART allows researchers to get draws from the posterior distribution of the predicted values. In Bayesian statistics, the posterior predictive distribution indicates the distribution of predicted data based on the data one has already seen \citep{gelman2014bayesian}. Hence, the posterior predictive distribution can be used to predict new data values and to draw inference around the distribution of these predicted values.

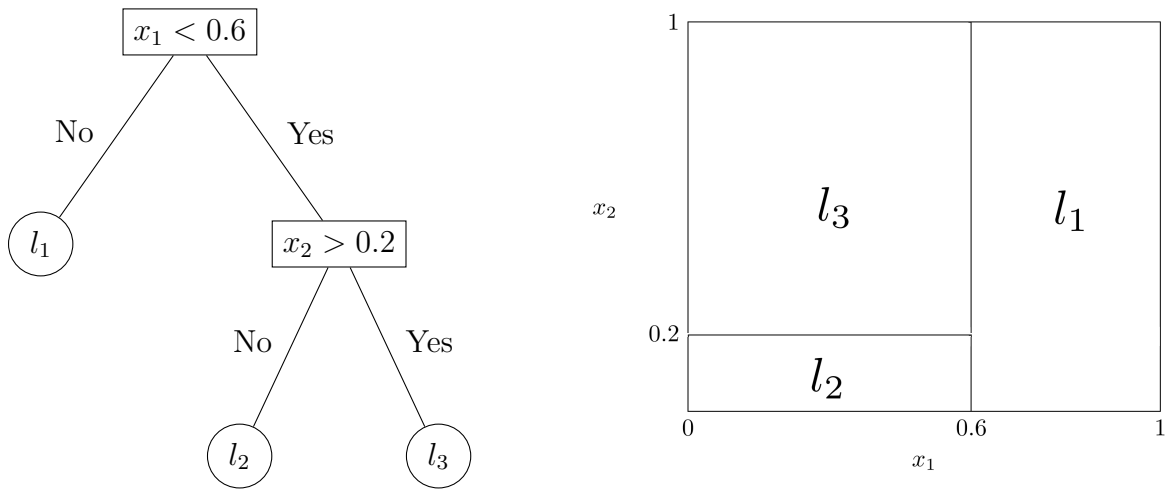
\begin{figure}
    \centering
    \begin{subfigure}[b]{0.48\textwidth}
        \centering
		\begin{tikzpicture}[level distance=80pt, sibling distance=50pt, edge from parent path={(\tikzparentnode) -- (\tikzchildnode)}]
		\tikzset{every tree node/.style={align=center}}
		\Tree [.\node[rectangle,draw]{$x_1<0.6$}; \edge node[auto=right,pos=.6]{No}; \node[circle,draw]{$l_1$}; \edge node[auto=left,pos=.6]{Yes};[.\node[rectangle,draw]{$x_2 > 0.2$}; \edge node[auto=right,pos=.6]{No}; \node[circle,draw]{$l_2$}; \edge node[auto=left,pos=.6]{Yes}; \node[circle,draw]{$l_3$};  ]]
		\end{tikzpicture}
    \end{subfigure}
    \begin{subfigure}[b]{0.48\textwidth}
    \includegraphics[width=\textwidth]{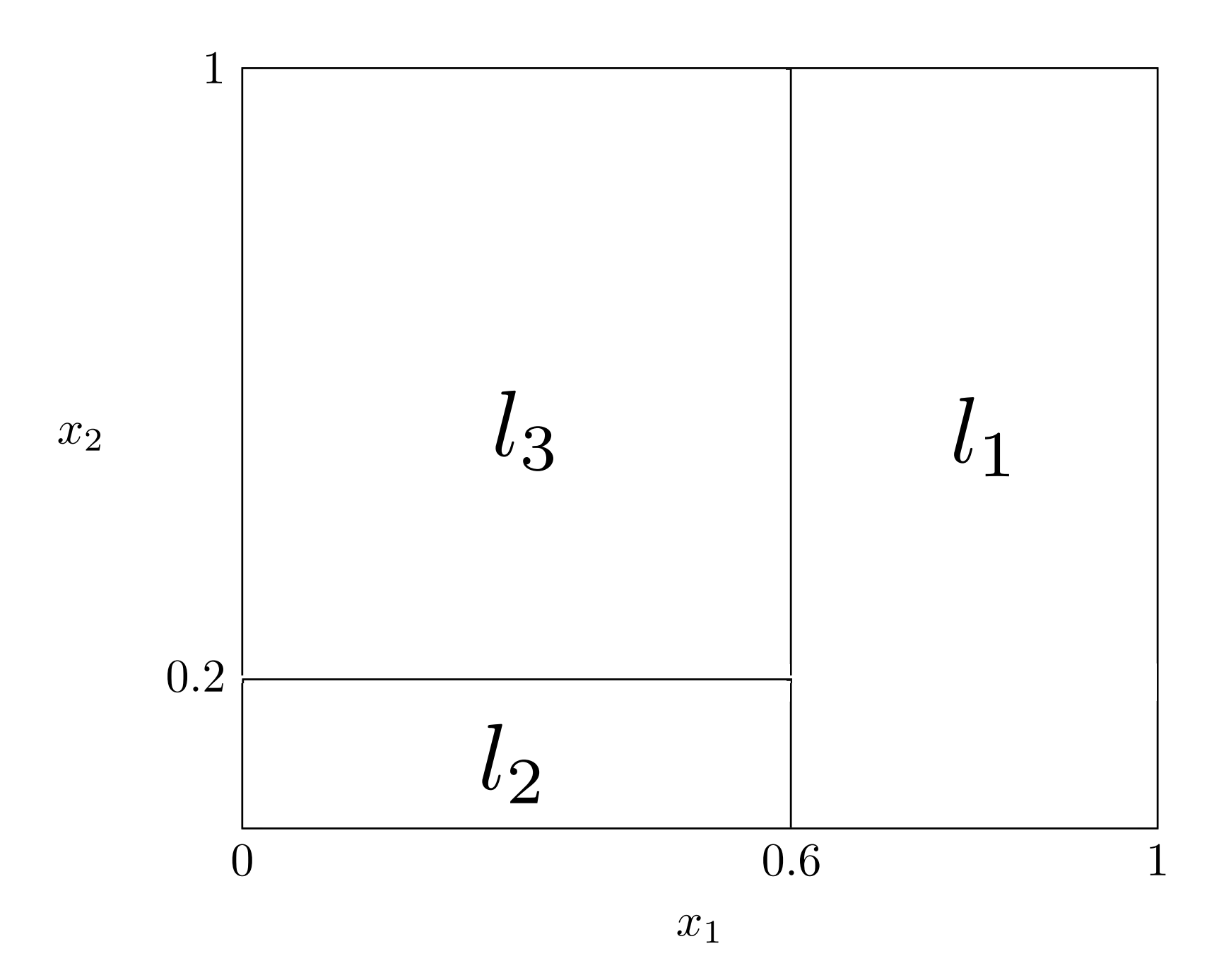}
    \end{subfigure}
    \caption{{\footnotesize(Left) An example of a binary tree. The internal nodes are labelled by their splitting rules and the terminal nodes are labelled with the corresponding parameters $l_i$.\\ (Right) The corresponding partition of the sample space.}}
    \label{fig:CART}
\end{figure}

BART finds its foundation in the CART algorithm. CART is a widely used algorithm for the construction of trees where each node is split into only two branches (i.e., binary trees). Figure \ref{fig:CART} illustrates how binary partitioning works in practice based on a simple case with just two predictors $x_1 \in [0,1]$ and $x_2 \in [0,1]$.

The accuracy of the predictions of binary trees can be dramatically improved by iteratively constructing the trees. BART is sum-of-trees ensemble algorithms, and its estimation approach relies on a fully Bayesian probability model \citep{kapelner2013bartmachine, JSSv070i04}. Let us define with $Y$ the outcome vector, with $y_i$ the outcome for a generic unit $i$, and with $\bX$ the $N \times P$ matrix of covariates or \textit{predictors} (where $P$ is the number of predictors), with $X_i$ the $P$-dimensional vector of predictors for $i$ (with $i=1,..., N)$.
Then, the BART model can be expressed as:
\begin{equation}\label{tree}  \centering 
    Y = f(\bX) + \epsilon \approx \mathcal{T}_1(\bX; \mathcal{D}_1, \mathcal{M}_1) + ... + \mathcal{T}_J(\bX; \mathcal{D}_j, \mathcal{M}_j) + \epsilon, \:\:\:\:\:\:\:\:\: \epsilon_i \sim \mathcal{N} (0, \sigma^2),
\end{equation}
where the $J$ distinct binary trees are denoted by $\mathcal{T}(\bX; \mathcal{D}_j, \mathcal{M}_j)$. $\mathcal{T}$ is a function that sorts each unit into one of the sets of $m_j$ terminal nodes, associated with mean parameters $\mathcal{M}_j = \{\mu_1, ..., \mu_{m_j}\}$, based on a set of decision rules, $\mathcal{D}_j$.
$\epsilon$ is an error term and is typically assumed to be independent, and identically normally distributed when the outcome is continuous \citep{chipman2010bart}.   
Using more compact notation we can re-write the BART model as:
\begin{equation} \label{formula:BART}
    Y = \sum_{j=1}^{J} \mathcal{T}_j(\bX; \mathcal{D}_j, \mathcal{M}_j) + \epsilon.
\end{equation}
The Bayesian component of the algorithm is incorporated in a set of three different priors on: (i) the structure of the trees, $\mathcal{D}_j$ (this prior is aimed at limiting the complexity of any single tree $\mathcal{T}$ and works as a regularization device); (ii) the distribution of the outcome in the nodes, $\mathcal{M}_j$ (this prior is aimed at shrinking the node predictions towards the center of the distribution of the response variable $Y$);  (iii) the error variance $\sigma^2$ (which bounds away $\sigma^2$ from very small values that would lead the algorithm to overfit the training data)\footnote{The choice of the priors and the derivation of the posterior distributions is discussed in depth by \cite{chipman2010bart} and \cite{JSSv070i04}. Namely, (i) the prior on the probability that a node will split at depth $k$ is $\beta(1+k)^{-\eta}$ where $\beta \in (0,1), \eta \in [0, \infty)$ (these hyper-parameters are generally chosen to be $\eta=2$ and $\beta = 0.95$); (ii) the prior on the probability distribution in the nodes is a normal distribution with zero mean: $\mathcal{N}(0, \sigma^2_q)$ where $\sigma_q = \sigma_0/\sqrt{q}$ and $\sigma_0$ can be used to calibrate the plausible range of the regression function; (iii) the prior on the error variance is $\sigma^2 \sim InvGamma(v/2, v\lambda/2)$ where $\lambda$ is determined from the data in a way that the BART will improve 90\% of the times the RMSE of an OLS model.}. The aim of these priors is to ``regularize" the algorithm, preventing single trees from dominating the overall fit of the model \citep{JSSv070i04}. 
 
These Bayesian tools give researchers the possibility to mitigate the overfitting problem of RFs and to tune the algorithm with prior knowledge. Moreover, BART has shown a consistently strong performance under ``default" model specifications. This is a highly valuable characteristic of BART as it reduces its dependence on the choice of parameters done by the researcher as well as the computational time and costs related to cross-validation.

Let us imagine a simple case in which we have a study sample ($\Omega = \{\bX,Y\})$ for which we observe the set of predictors $\bX$, but the vector of  outcomes is observed just for a subsample of the study population, $Y^{obs}$. Hence, we can indicate with $Y^{mis}$ the vector of missing outcomes in the study sample, where $Y=Y^{mis}\cup Y^{obs}$. To obtain predictions for these values, the first step is to build a BART model as in \eqref{formula:BART} regressing $Y^{obs}$ on the observed matrix of predictors. To perform the imputation for $Y^{mis}$, BART collects a sample from the posterior distribution of $\theta = \{\sigma^2, \mathcal{D}_j, \mathcal{M}_j\}$, $p(\theta|Y^{obs})$ using the Bayesian backfitting algorithm proposed by \cite{chipman2010bart}. An imputed value for $\hat{Y}^{mis}$ can be obtained by sampling from the posterior predictive distribution:
\begin{equation} \label{eq:missing}
    p(Y^{mis}|Y^{obs}) = \int p(Y^{mis}|Y^{obs}, \theta) \cdot p(\theta|Y^{obs}) \: d\theta.
\end{equation}

From \eqref{eq:missing} we obtain draws from the posterior predictive distribution of $Y^{mis}$. Then, we can use standard techniques employed for outlier detection, such as the ones proposed by \cite{miller1991reaction} (i.e., $K$ standard deviation from the mean of the distribution) and \cite{leys2013detecting} (i.e., $K$ absolute deviations from the median), to detect the predictions with values that are consistently further away from the mean or the median of the predicted distribution, respectively. The nice feature of such techniques, is that they can be manually tuned in order to include units with more or less extreme predicted values.\footnote{In case of multivariate predictions, one could implement data driven methodologies for outliers detection such as the Isolation Forest \citep{liu2012isolation}.}

These analyses can be highly relevant for policy-makers to spot observations with more or less extreme predicted outcome values. Indeed, it could be the case that policy-makers are interested in targeting their intervention to a specific subpopulation, based on the values of an unobserved, yet predictable, outcome (e.g., provide additional learning material to more vulnerable students in a region where financial literacy is not observed). Moreover, information on the factors related to these ``high or low levels" of the predicted outcome are not only useful to identify potential subgroups but also to reveal which factors are associated with specific levels or values of the outcome. 

\subsection{Sensitivity of predictions analysis} \label{subsec:sensitivity}

The ability of machine learning techniques to provide accurate and \textit{robust} predictions is key. As predictions always hold a degree of uncertainty, we argue that there is a need for specific techniques that enable researchers and practitioners to assess their \textit{sensitivity}. This is particularly true as the usage of machine learning for prediction policy problems has been increasing over the last few years \citep{kleinberg2015prediction, mullainathan2017machine}. 

The main objective of the \textit{sensitivity of predictions analysis} introduced here is to assess how the omission of an unobserved predictor could affect both the forecasts of the model and its performance. Indeed, we argue that a desirable property of any predictive model is to be robust to potentially unobserved predictors. In a predictive scenario, we can define \textit{robustness} as the model's ability to provide stable predictions and a stable performance over the inclusion of additional variables in the model. Robustness of predictions is a desirable property as volatile predictions may be completely overturned by the inclusion of an additional predictor in the model. Stability of predictions  per-se is not necessarily a desideratum: a model that is completely unable to properly predict a certain outcome may be stable but highly inaccurate. We argue that the stability of the model is strictly connected with its performance and depends on the quality of the additional predictor included. As one includes a new predictor with a high explanatory power in a poorly performing model (i.e., a model with highly inaccurate predictions), the model will depict a boost in its predictive power driven by an increase in the precision of its predictions. Hence, it is intuitive to see how such a sensitivity of prediction analysis is important for at least two reasons. First, it can show how much an additional predictor could boost the performance of the model, indicating how much the model can be improved by taking into account additional, potentially unobserved predictors. Second, it reveals the variability in the predictions induced by novel predictors: the more stable they are, given a high predictive performance, the more the model can be trusted for predictive policy tasks.

Such a sensitivity of prediction analysis is directly inspired by its causal counterpart \citep{rosenbaum1983assessing, ichino2008temporary}. In particular, \cite{ichino2008temporary} propose a simulation-based sensitivity analysis to assess the robustness of causal estimands to failures in the unconfoundedness assumption (i.e., independence between the treatment and the potential outcomes conditional on the observed covariates). The authors build on the works by \cite{rosenbaum1983central} and \cite{rosenbaum1983assessing} and simulate an unobserved confounder that is associated with both the potential outcome and the treatment. Then, they introduce this additional confounder in the model for the estimation of the Average Treatment effect on the Treated (ATT). They repeat the estimation procedure many times using a different set of values for the simulated confounder and compare these estimates with the original estimate of the ATT obtained without the inclusion of the simulated confounder. Such comparison informs scholars on how robust the original causal estimands are to different configurations used for the construction of the confounder (and, in turn, different levels of violations of the unconfoundedness assumption).

While departing from a similar intuition, the sensitivity analysis that we implement here is fundamentally different because we deal with a prediction setting rather than a causal inference one. In fact, we depart from the conceptual meaning of causal sensitivity analyses as tools to assess the robustness of an estimators to an underlying and untestable assumption. In this sense, the sensitivity analysis that we introduce in this paper is not aimed at testing any identification assumption. Conversely, our aim is to provide a direct assessment of the quality of the model and its predictions. The added value of the proposed approach is in those situations where one has a well-performing predictive model and wants to assess whether or not it's worthwhile to continue the search for new predictors.

The sensitivity of prediction analysis is performed by generating a \textit{synthetic} predictor and checking if (and how) its inclusion in the set of observed predictors changes the model's predictions and, in turn, its performance. As we want this predictor to have a good explanatory power with respect to the outcome vector $Y$, we will design it to have a high correlation with the outcome and to be uncorrelated with all the other predictors in the model. 

Before going into the nuts and bolts on how to construct the synthetic variable, let us discuss the implications of constructing this synthetic variable in such a way. While a high correlation with the outcome and a zero correlation with the observed predictors guarantees a high explanatory power -- i.e., a large share of the variation in the outcome being explained by the synthetic variable, controlling for the other predictors -- it does not guarantee that the synthetic variable is also the one with the highest predictive power. Indeed, it may be the case that in a non-linear setting, the interaction between two variables with lower explanatory power and the outcome could carry more information than a predictor with a higher explanatory power than these two individual variables. However, such interaction patterns between the predictors and the outcome are hidden in the data (in some way, the ultimate goal of supervised learning methodologies is to discover such underlying patterns), and is not possible to reproduce them in any meaningful way. Hence, constructing a variable with a high predictive power, while it may not lead to the best absolute model performance, is (i) feasible (while it may not be possible to construct a synthetic variable with the highest predictive power due to the complex, non-linear patterns of interactions between the predictors and the outcome) and; (ii) meaningful: the explanatory power of a variable is relevant also in a prediction setting (especially when the ultimate users are the policy-makers).

\subsubsection{Building the synthetic predictor}
Let us define with $R$ a $N$-dimensional vector for the synthetic predictor. We start from sampling $N$ independent realizations from a normally distributed continuous variable: $\tilde{R} \sim \mathcal{N}(0,1)$. The distribution of this variable mimics the distribution of the synthetic predictor and can be directly changed by the researcher. We let $\tilde{\mathbf{R}}$ be the $N \times (P+2)$ matrix that stacks the output vector $Y$, $\tilde{R}$ and the matrix of observed variables $\mathbf{X}$ such that:
\begin{equation}
    \tilde{\mathbf{R}}_{(N\times (P+2))} = \big[Y_{(N \times 1)} \:\: \tilde{R}_{(N\times 1)} \:\: \mathbf{X}_{(N\times P)} \big].
\end{equation}
We define with $\hat{\mathbf{K}}$ the estimated correlation matrix of $\tilde{\mathbf{R}}$,
\begin{equation}
\hat{\mathbf{K}}_{((P+2)\times (P+2))} = 
\begin{pmatrix}
    1 & \hat{\rho}(Y,\tilde{R}) &  \ldots & \hat{\rho}(Y,X_p) \\
    \hat{\rho}(\tilde{R},Y) & 1 &  \ldots &  \hat{\rho}(\tilde{R}, X_p) \\
    \vdots & \vdots & \ddots & \vdots \\
    \hat{\rho}(X_p, Y) & \hat{\rho}(X_p, \tilde{R}) & \ldots & 1
\end{pmatrix}
\end{equation}
and with $\tilde{\mathbf{\Sigma}}= diag(\hat{\sigma}_{\tilde{Y}}^2, \hat{\sigma}_{\tilde{R}}^2, \hat{\sigma}_{X_1}^2 ..., \hat{\sigma}_{X_p}^2)$ the $(P+2) \times (P+2)$ diagonal matrix of estimated variances, where $\hat{\sigma}_{\tilde{Y}}^2$ is the estimated variance of the output, $\hat{\sigma}_{\tilde{R}}^2$ is the estimated variance of $\tilde{R}$ and $\hat{\sigma}_{\tilde{X_j}}^2$ is the estimated variance of the $j$-th variable in $\bX$. Moreover, we let $\mathbf{C}$  be the symmetric, idempotent centering $N \times N$ matrix:
\begin{equation}
    \mathbf{C} = \mathbf{I} - \frac{1}{N} \mathbf{1}
\end{equation}
where $ \mathbf{I}$ is the identity matrix and  $\mathbf{1}$ is a $N \times N$ matrix of 1s. We center and scale the $\mathbf{\tilde{R}}$ matrix as follows:
\begin{equation}
    \dot{\mathbf{R}} =  \mathbf{C} \tilde{\mathbf{R}} \tilde{\mathbf{\Sigma}}^{-1}
\end{equation}
where $\dot{\mathbf{R}}$ is a multivariate normal matrix. This is similar to the type of scaling used, for example, in standard normal variate correction, and corresponds to a variation of the whitening (or sphering) transformation.
Then, we introduce no-correlation within the columns in $\tilde{\mathbf{R}}$ as follows multiplying 
\begin{equation} \label{fomula:uncorrelated}
    \mathbf{L}^{-1}\dot{\mathbf{R}}
\end{equation}
where $\mathbf{L}$ is a lower triangular matrix obtained through  the Cholesky decomposition of $\tilde{\mathbf{K}}$: $\tilde{\mathbf{K}}=\mathbf{L}\mathbf{L}^T$.
The matrix that we obtain in \eqref{fomula:uncorrelated} is a set of realizations of uncorrelated multivariate normal random variables. At this point, we can reintroduce the correlation between each column as follows:
\begin{equation}
   \mathbf{U}\mathbf{L}^{-1}\dot{\mathbf{R}}  
\end{equation}
where $\mathbf{L}$ is obtained thought as the Cholesky decomposition of $\mathbf{K}$\footnote{This is possible as far as  $\mathbf{K}$ is a symmetric, positive-definite matrix. In the case of a semi-positive definite matrix, one can employ a global shift $\alpha > 0$ and then compute a Cholesky decomposition of the scaled and shifted matrix: 
    \begin{equation*}
      \mathbf{K}=\mathbf{S}\mathbf{U}\mathbf{U}^T\mathbf{S} + \alpha \mathbf{I} 
    \end{equation*}
where $\mathbf{S}$ is diagonal scaling matrix. This shift is known as Tikhonov regularization.}, $\mathbf{K}=\mathbf{U} \mathbf{U}^T$, and $\mathbf{K}$ is a newly defined correlation matrix:
\begin{equation}
\mathbf{K}_{((P+2)\times (P+2))} = 
\begin{pmatrix}
    1 & \widetilde{\rho}(Y,R) &  \ldots & \hat{\rho}(Y,X_p) \\
    \widetilde{\rho}(R,Y) & 1 &  \ldots & \widetilde{\rho}(R, X_p) \\
    \vdots & \vdots & \ddots & \vdots \\
    \hat{\rho}(X_p, Y) & \widetilde{\rho}(X_p, R) & \ldots & 1
\end{pmatrix}
\end{equation}
where $\widetilde{\rho}(Y,R)$ is the correlation between the synthetic predictor $R$ and the outcome, that is set directly by the researcher, and $\widetilde{\rho}(X_j,R)$ is the correlation between the synthetic predictor the $j-$th regressor. 
These correlations are set to be equal to zero in our first setting:
\begin{equation}
\mathbf{K}_{((P+2)\times (P+2))} = 
\begin{pmatrix}
    1 & \widetilde{\rho}(Y,R) &  \ldots & \hat{\rho}(Y,X_p) \\
    \widetilde{\rho}(R,Y) & 1 &  \ldots & 0 \\
    \vdots & \vdots & \ddots & \vdots \\
    \hat{\rho}(X_p, Y) & 0 & \ldots & 1
\end{pmatrix}
\end{equation}
In a further step, we make this algorithm robust by introducing correlations between the synthetic variable and the best predictor, to closely mimic the case of real-world applications where predictors are correlated. The choice of the correlations pattern is left to the researcher, and can be informed by previous analyses on the relative importance and correlation of the predictors.

 We can now rework the model in \eqref{formula:BART} to include the set of stack predictors $\mathbf{G}$:
\begin{equation}\label{eq:bart_synth}
     \sum_{l=1}^{L} \mathcal{T}_l(\mathbf{G}; \mathcal{D}_l, \mathcal{M}_l) + \psi
\end{equation}
where $\mathbf{G}$ is:
\begin{equation}
    \mathbf{G}_{(N \times (P+1))} = \big[{R}_{(N\times 1)} \:\: \mathbf{X}_{(N\times P)} \big].
\end{equation}
Lastly, we scale the data back to their original distribution:
\begin{equation}
    \mathbf{R}=\mathbf{C}^{-1}\mathbf{U}\mathbf{L}^{-1}\dot{\mathbf{R}}\tilde{\mathbf{\Sigma}}
\end{equation}
and ``extract" the newly generated synthetic predictor $R$ from the $\mathbf{R}$ matrix.  In this specific case, the synthetic predictor corresponds to the second column of the matrix $\mathbf{R}$. However, its relative position depends on how the matrix $\tilde{\mathbf{R}}$ is generated.

Finally, we compare the predictions of the new \textit{augmented} model in (\ref{eq:bart_synth}) with the ones of the \textit{original} BART model in (\ref{formula:BART}).
To do so, we run the two models on $b$ bootstrapped samples $b=1,...,B$ to get a set of predicted values at each iteration:
\begin{eqnarray}
	\sum_{j=1}^{J} \mathcal{T}_{j,b}(\mathbf{X}; \mathcal{D}_j, \mathcal{M}_j) + \epsilon &=& \hat{Y}_{b}(\bX); \\
    \sum_{l=1}^{L} \mathcal{T}_{l,b}(\mathbf{G}; \mathcal{D}_l, \mathcal{M}_l) + \psi &=& \hat{Y}_{b}(\mathbf{G}).
\end{eqnarray}
Once we get the unit level predictions for these models at each iteration step $b$ ($\hat{y}_{b}(x_i)$ and $\hat{y}_{b}(g_i)$, respectively), we can run a series of comparisons between both the distribution of the predictions and the performance of augmented model as compared to the original model.  To assess the statistical difference between the unit level predictions' distributions we assess the difference between the mean of the unit level prediction of the augmented model, $\bar{\hat{y}}(x_i)$, and the mean of the unit level predictions of the original model, $\bar{\hat{y}}(r_i)$:
\begin{equation*}
    \bar{\hat{y}}(x_i) = {1 \over B} \sum_{b=1}^{B} \hat{y}_{b}(\mathbf{X}=x_i) \:\:\:\:\:\: \text{and} \:\:\:\:\:\: \bar{\hat{y}}(r_i) = {1 \over B} \sum_{b=1}^{B} \hat{y}_{b}(\mathbf{G}=g_i).
\end{equation*}
To assess whether or not there is an improvement in the performance of the augmented model, we compare the RMSE of prediction and the adjusted $R^2$ of the original and augmented models. As the correlation coefficients between the synthetic predictor and the outcome are subjectively defined, these comparisons can be drawns for different correlation settings. The more similar the unit level predictions and performance of the original and the augmented model, the more one can argue that the available signal is enough for stable and precise predictions.

The table Algorithm \ref{alg:dis} summarizes the algorithm introduced in this Subsection and its main steps.\footnote{Here, we introduce our algorithm in its most general version. However, running $B$ times the machine learning model at hand can be computationally intensive, especially for less scalable machine learning methodologies. In the case of BART, to reduce such computational burden, one can directly sample the unit level predictions from the posterior predictive distribution. For the RF algorithm, we argue that researchers can invoke the statistical properties introduced by \cite{wager2018estimation} to construct reliable confidence intervals and statistical tests for each unit level prediction.}

\vspace{1cm}
\begin{algorithm}
\caption{Overview of the sensitivity analysis \label{alg:dis}}
\vspace{0.25cm}
The steps of the algorithm:
	\begin{enumerate}
			\item Create a new matrix of predictors $\mathbf{G}$ stacking the observed predictors $\bX$ and a synthetic predictor $R$ correlated with the outcome $Y$ and uncorrelated to all the variables in $\bX$;
			\item Generate two models, one including only the observed predictors $\bX$ (\textit{original model}) and one including the observed predictors and the synthetic predictor $\mathbf{G}$ (\textit{augmented model});
			\item Run the two models on $b$ bootstrapped samples ($b=1,...,B$) to get a set of predicted values at each iteration for $\hat{Y}_{b}(\bX)$ and 
            $\hat{Y}_{b}(\mathbf{G})$;
			\item Compare the original and the augmented model with respect to:
			\begin{enumerate}
			    \item Their unit level predictions;
			    \item Their performance.
			\end{enumerate}
        \end{enumerate}	
	\end{algorithm}

\section{Data} \label{sec:data}

To illustrate the proposed methodologies for prediction and sensitivity, we apply them to the 2015 data of the OECD's Program for International Student Assessment (PISA) for Belgium. PISA is a worldwide study aimed at evaluating educational systems by measuring students scholastic performance on reading, mathematics, and science. These assessments are intended to provide comparable data that can, in turn, enable countries to improve their education policies and outcomes.

The structure of the Belgian PISA 2015 data offers an interesting and relevant example of forecasting missing scores on a standardised student assessment. All regions of the country participated in the general assessment of PISA, while only selected regions participated in the financial literacy assessment. The Belgian federal state has three regions: Flanders, Brussels and Wallonia. While the former two participated in both the general and financial literacy assessment, the latter participated only in the general assessment.\footnote{To simplify, we use Flanders to denote both the Flemish and Bruxelles region.} This structure of the data allows us to use a common set of predictors for all regions from the general assessment in order to predict financial literacy outcomes for students in the region that did not participate in this part of PISA.\footnote{In the eight OECD countries participating in the PISA financial literacy assessment 2015, Belgium and Canada are the ones with the most remarkable differences in regional assessments.}

We choose the Belgian case to illustrate the forecasting methodology, since the regions of the country have many similarities and common policies that justify the assumption of similar underlying relationships between the predictors and the outcome variable. At the same time, the Belgian regions differ in population characteristics which suggests that the missing scores in Wallonia are unlikely to be identical to those in Flanders. In the next two Subsections, we will describe the set of predictors (Subsection \ref{subsec:predictors}) and the outcome (Subsection \ref{subsec:outcome}) in more detail.

\subsection{Predictors}\label{subsec:predictors}

In contrast to regression analyses, in which multicollinearity of regressors needs to be avoided, a large number of (potentially correlated) predictors can be used for forecasting in machine learning \citep{vaughan2005using, makridakis2008forecasting, shmueli2010explain}. We therefore select a broad set of predictors from the general assessment of PISA 2015. Table \ref{tab:variables} provides an overview of the variables used as predictors. The set of predictors includes: (i) students' background characteristics; (ii) proxies of the socioeconomic status of students; (iii) indicators of student achievement and attitudes; (iv) school characteristics.

To account for the students' background, we include gender, grade, age, language and study track. Existing studies commonly find FLS to be highly correlated in particular with math  and reading performance \citep[e.g.,][]{Mancebon2019,Riitsalu2016}. In the case of Flanders, the correlation of FLS with math  and reading scores amounts to 0.8, which is slightly higher than the OECD average of 0.75 \citep{UniversiteitGent2017}. As such, we include the PISA math  score, as well as the reading score and additional variables related to student achievement, such as grade repetition, study time and instruction time. Given that personality traits and attitudes have been found to matter for financial literacy as well \citep{Longobardi2018,Pesando2018}, we also include students' anxiety and motivation levels recorded using standard tests in the PISA survey. 

Another major factor associated with FLS is the family background, such as parental characteristics, language or immigration background \citep[e.g.,][]{Mancebon2019,Gramatki2017}. To approximate the socioeconomic status of a student, we use the indicators of educational and economic possessions of the family, the number of books in the home, the immigration background of the student, mother's and father's education and job, as well as a variable capturing perceived parental emotional support.

Finally, studies of PISA financial literacy outcomes commonly include variables for school level variables \citep[e.g.,][]{Pesando2018,Cordero2018}. We include a number of school characteristics from the questionnaire for school principals, such as the school's location, size and autonomy. As indicators of teaching quality, we use the student-teacher ratio, class size and teacher professional development. We also use school level indicators of socioeconomic status, such as the share of students with a different home language, special needs or a socioeconomically disadvantaged home, the number of available computers and shortage of educational material.

Figure \ref{fig:missingness} in Appendix \ref{subsec:app-data} shows the missing observations in the variables used in the analysis. Apart from the FLS, which are, as described above, only available for Flemish students, no clear patterns of missingness appear across the predictors. We can therefore assume the observations to be Missing-Completely-at-Random \citep{little2019statistical} and we proceed with multiple imputations using the Fully Conditional Specification (FCS) developed by \cite{buuren2010mice} and implemented in the \texttt{R} package \texttt{MICE}.

\begin{figure}[H]
    \centering
    \begin{subfigure}[b]{0.48\textwidth}
    \includegraphics[width=\textwidth]{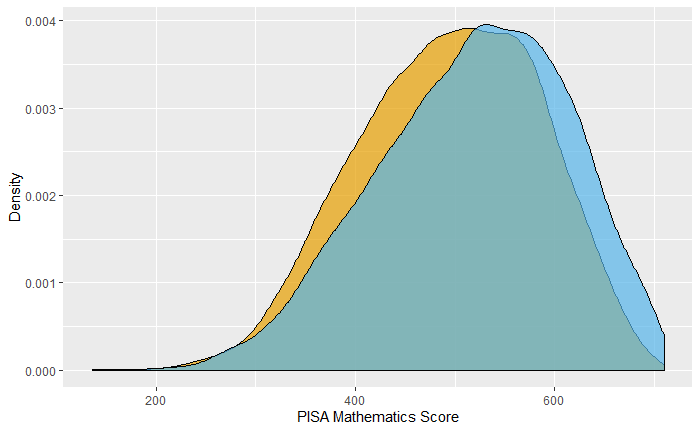}
    \end{subfigure}
    \begin{subfigure}[b]{0.48\textwidth}
    \includegraphics[width=\textwidth]{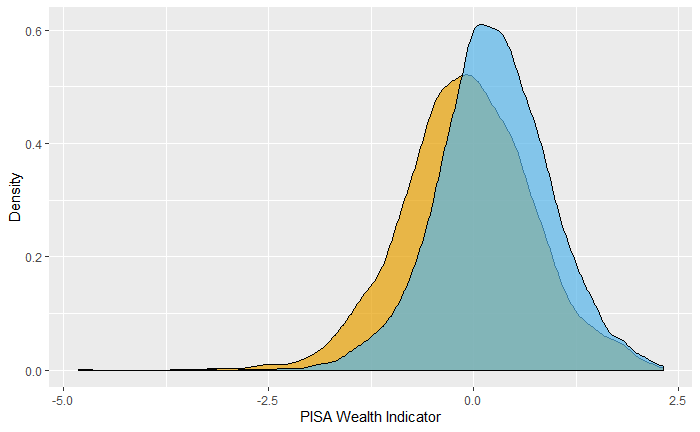}
    \end{subfigure}
    \caption{{\footnotesize(Left) Distribution of PISA math  Scores in Flanders (blue) and Wallonia (yellow).\\ (Right) Distribution of PISA Wealth Index in Flanders (blue) and Wallonia (yellow).}}
    \label{fig:mathscore}
\end{figure}

As shown in Table \ref{tablesummarystats} in Appendix \ref{subsec:app-data}, a series of T-tests reveal significant differences in means on background characteristics of students in Flanders and Wallonia. The Belgian sample is balanced across regions in terms of gender, age and parents' characteristics. However, potentially important variables with respect to financial literacy, such as math  and reading scores, as well as the socioeconomic status, are significantly different in the two regional samples. Figure \ref{fig:mathscore} shows that the distribution of math scores in Flanders and Wallonia overlaps; however the distribution is shifted to the right for Flanders compared to Wallonia. A similar pattern is observed regarding the socioeconomic status of students in the two samples, as approximated by the PISA wealth indicator which summarises economic possessions of the family. Figure \ref{fig:mathscore} shows that, similarly, the distribution of wealth overlaps, but is slightly shifted to the right in Flanders compared to Wallonia. 

Even though there are obvious differences in the univariate distributions of the predictors, we find that their multivariate distributions largely overlap. In particular, we develop an overlap score aimed at assessing if there are regions of the features' space where there is no overlap between the multivariate distribution of the predictors in Flanders and Wallonia. Additional details on the proposed methodology can be found in Appendix \ref{subsec:overlap}. Since we find that the multivariate distributions of the predictors largely overlap for the two regions, this hints at the fact that our predictive model will not extensively rely on extrapolation to extend its predictions from Flanders to Wallonia.

\subsection{Outcome Variable}\label{subsec:outcome}

The FLS in PISA 2015 are reported as a set of ten plausible values.\footnote{For each participating country or economy, 33\% of the students who completed the general PISA assessment were tested on financial literacy \citep{OECD2017b}. The plausible values were then constructed for all students participating in the general PISA based on item response theory and latent regression. In the case of Belgium, this was only done for Flanders. As a results, the plausible values are available for all participating students in the Flemish region, but missing for students from Wallonia. In the following, we use the plausible value $PV1FLIT$ for the analysis, since any bias caused by using a single plausible value instead of all ten simultaneously is arguably negligible \citep{Gramatki2017}.} The financial literacy score is based on a test of financial knowledge with 43 items in four content categories: (i) money and transactions; (ii) risk and reward; (iii) planning and managing finances; and (iv) the financial landscape \citep{OECD2017a}. 

In the case of Belgium, there are 9,651 observations on the general assessment from all Belgian regions, but the FLS are only available for the 5,675 students from Flanders, while the FLS of the 3,976 students from Wallonia are missing. Flanders thus represents our ``study" population in which we observe the FLS $Y^{obs}$. Wallonia represents the ``target" population for which we predict the missing FLS, $Y^{mis}$, based on the common set of predictors from the general assessment $\bX$. Table \ref{tableoutcome} in Appendix \ref{subsec:app-data} provides summary statistics of the outcome variable. On average, Flemish students score 541 points, which corresponds to the PISA proficiency level 3 out of 5 levels. Figure \ref{fig:outcome} in Appendix \ref{subsec:app-data} shows the distribution of FLS in the Flemish data. 12\% of Flemish students fail to reach the baseline level 2 of 400 points or more which the OECD defines as the level necessary to participate in society \citep{OECD2017a}. Figure \ref{fig:PISAlevels} in Appendix \ref{subsec:app-data} provides an overview of the required knowledge corresponding to the five PISA proficiency levels.

\section{Results} \label{sec:results}

In this Section, we discuss the performance of BART in the prediction task at hand. Then, we provide the results and some descriptive evidence regarding the observation with posterior predicted probability lower than the average (vulnerable students). Finally, we depict and discuss the results from the sensitivity analysis. 

\subsection{Results for BART predictions}

Many recent studies have shown that BART performs in an excellent way in various predictive tasks across different scenarios \citep{murray2017log,linero2018bayesian1, linero2018bayesian2, hernandez2018bayesian, bargagli2020machine, bargagli2020supervised}. In our particular scenario, we assess the performance of BART through a 10-folds cross-validation, and compare it with the performance of the random forest (RF) algorithm \cite{breiman2001random}.  A generic $k$-folds cross-validation technique consists of dividing the overall sample (in our specific case we draw this sample from the set of observations for which the outcome variable is present) into $k$ different subsamples. $k-1$ subsamples are used to train the machine learning technique. Then its performance is assessed on the subsample that was left out.  We decide to compare the performance of our algorithm with the one of the RF algorithm, one of the most widely used algorithms for prediction in social sciences \citep{athey2018impact, bargagli2020supervised}. We observe that BART performs equally or better than the random forest algorithm in the predictive task at hand. Table \ref{table:performance} depicts the comparative results of the two algorithms with respect to their root-mean-squared-error (RMSE) and mean-absolute-error (MAE) of prediction and their $R^2$. BART demonstrates a very good performance (as well as a series of other qualities that were described briefly in Section \ref{sec:methodology}), as it outperforms RF with respect to all the selected performance measures (i.e., smaller RMSE and MAE, higher $R^2$). Hence, we use this technique to predict the missing FLS for students in Wallonia.

\begin{center}
\input{Tables/performance.tex}
\end{center}

\begin{figure}
    \centering
    \includegraphics[width=0.9\textwidth]{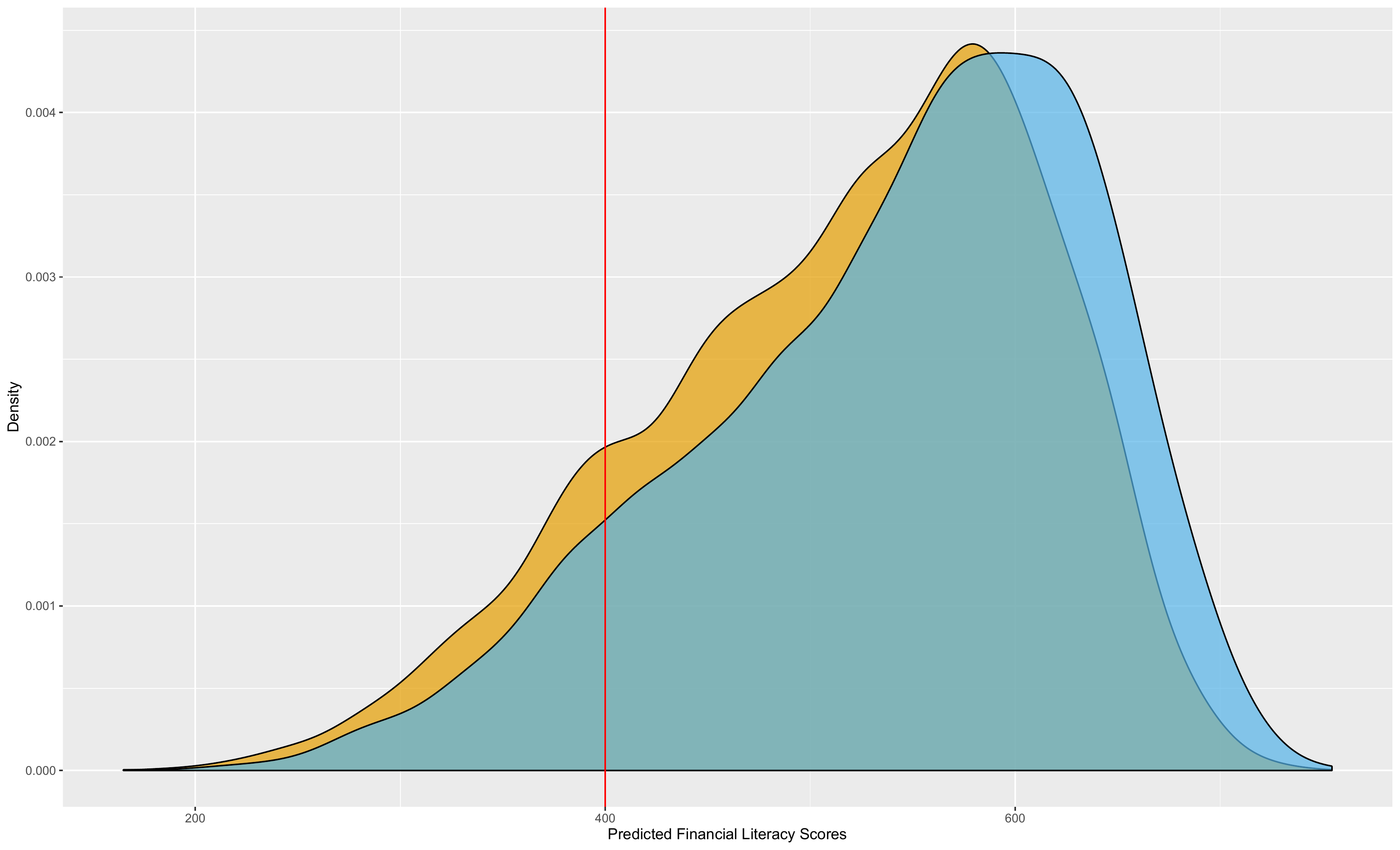}
    \caption{Predicted FLS for Flanders (light blue) and Wallonia (orange). The red line indicates the threshold of the baseline level of proficiency in financial literacy. OECD suggests that students above this threshold of 400 points have financial literacy levels that are sufficient to participate in society \citep{OECD2017a}.}
    \label{fig:predictions}
\end{figure}

Figure \ref{fig:predictions} depicts the posterior predicted values for FLS for students in Flanders (light blue) and Wallonia (orange), while Table \ref{table:fls} reports summary statistics for the same predictions. From Table \ref{table:fls} we observe that the mean posterior predictive FLS for students in Flanders is higher 
than for those in Wallonia, 541.4 compared to 516.6. Considering the minimum and maximum values for Flanders and Wallonia, we note that the scores of the students in Flanders are more centered around the mean in Wallonia compared to Flanders. The same conclusion can be drawn from Figure \ref{fig:predictions} where the peak of the light blue area (Flanders) is to the right of the peak of the orange area (Wallonia).
As OECD provide interpretable difference between the FLS, we can interpret the different results for Flanders and Wallonia. In particular, based on OECD computations, a difference of 40 points equals an entire school year of learning. Hence, based on this, we can state that, on average, students in Flanders are roughly half a school-year ahead of students in Wallonia with respect to their standardized knowledge in financial literacy.

\begin{center}
\input{Tables/stats_outcome.tex}
\end{center}

The quality of the predictions of FLS in Wallonia is not directly testable as the outcome is unobserved in this region. However, the quality of these predictions can be potentially very similar to the one obtained through cross-validation if the underlying assumption that the `technology' and the efficiency to transform teaching and environmental inputs into learning output is comparable between Flanders and Wallonia. As we argued in Section \ref{sec:data},this assumption is likely to hold. 

To investigate these assumption thoroughly, we run a series of robustness checks using different outcome variables. Given the high correlation between the outcome and math and reading scores, and given the standardized way all these outcome are recorded, the most straightforward way to implement this analysis is to use these two variable as outcomes and assess whether the BART model build using Flemish data performs well when tested on Walloon data. The results for these additional analyses are depicted in Appendix \ref{subsec:robustness}. We find no relevant differences between the performance of the model built using the Flemish data both for training and testing and the one built using Flemish data for training and Walloon data for testing. This robustness check seems to confirm the fact that the `technology' and the efficiency in Flanders and Wallonia are comparable.

\subsection{Results for vulnerable students detection}

Policy-makers often want to target only those who are really in need for an intervention as they are often facing budget constraints that prevent them from enacting policies for all constituents. These groups can be identified, in cases where the values of a certain outcome of interest are observed, by simply looking at the outcome's distribution. However, it is often the case that such outcomes may be hidden or not observed for a part of the study population. In such scenarios, machine learning predictions may furnish a valuable tool to policy makers. In the case of our application, we face a prediction problem: imagine that policy makers want to provide additional learning material to more vulnerable students in a region where FLS are not observed. Machine learning can provide a useful tool to draw such predictions.

In order to detect the most vulnerable students (i.e., those with low predicted FLS), we use the procedure introduced in Section \ref{subsec:bart}. Uncovering subpopulations of students that differ in terms of the distribution of a certain outcome is central to boost schools' and teachers' effectiveness and various strategies have been applied to this task \citep{masci2019semiparametric}. Interestingly, Bayesian machine learning has been applied to discover the relationship between financial literacy and self-reported economic outcomes \citep{puelz2020financial}. However, to the extent of our knowledge, this is the first time that a supervised machine learning technique is used to predict students' FLS.

In a previous study on OECD data, \cite{DeBeckker2019} identify four groups of adults with different financial literacy levels through an unsupervised machine learning algorithm (i.e., k-means). The advantage of the use of machine learning is clear as it allows to partition a large heterogeneous group of people into subgroups according to their FLS. However, while simple unsupervised machine learning algorithms such as k-means are useful to get a first impression of the different subgroups and their socioeconomic characteristics, they do not provide immediate information on the robustness of the outcomes. Moreover, these simple algorithms are not capable of making predictions for out-of-sample regions. This is particularly relevant when dealing with standardized tests, such as PISA, which often only cover subpopulations of a country, such as specific regions.

First, we train BART on the sample of Flemish students for which we have data on FLS. Next, we predict the FLS for both the sample of students in Flanders and Wallonia. After that, we compare the posterior predicted values for each student with the mean of the predicted posterior values. Finally, we detect the students for which these values are lying outside the credibility intervals for the mean values. Here, we define outliers as those observations with predicted FLS values smaller than two standard deviations from the mean following \cite{miller1991reaction}. In Appendix \ref{appendix:outliers}, we check the results we would obtain when defining the outliers as observations with predicted FLS two absolute deviations smaller than the median as suggested by \cite{leys2013detecting}.\footnote{Such results are not significantly different from the ones that are depicted in this Section. Thus, we argue that our results are robust to different definitions of outlying predictions.}

To get a better understanding of which variables best explain low predicted FLS, we generate a dummy variable that assumes value 1 if the student has a low predicted FLS and 0 otherwise. Finally, we built a series of conditional inference trees \citep{hothorn2006unbiased} to have a descriptive sense of which are the drivers of low financial literacy.  This fit-the-fit procedure has been widely applied in recent years to obtain interpretable results for the main drivers in the heterogeneity of an outcome \citep{lee2018discovering,bargagli2019heterogeneous,bargagli2020causal,lee2020causal}.

\begin{figure}[H]
    \centering
    \includegraphics[width=1\textwidth]{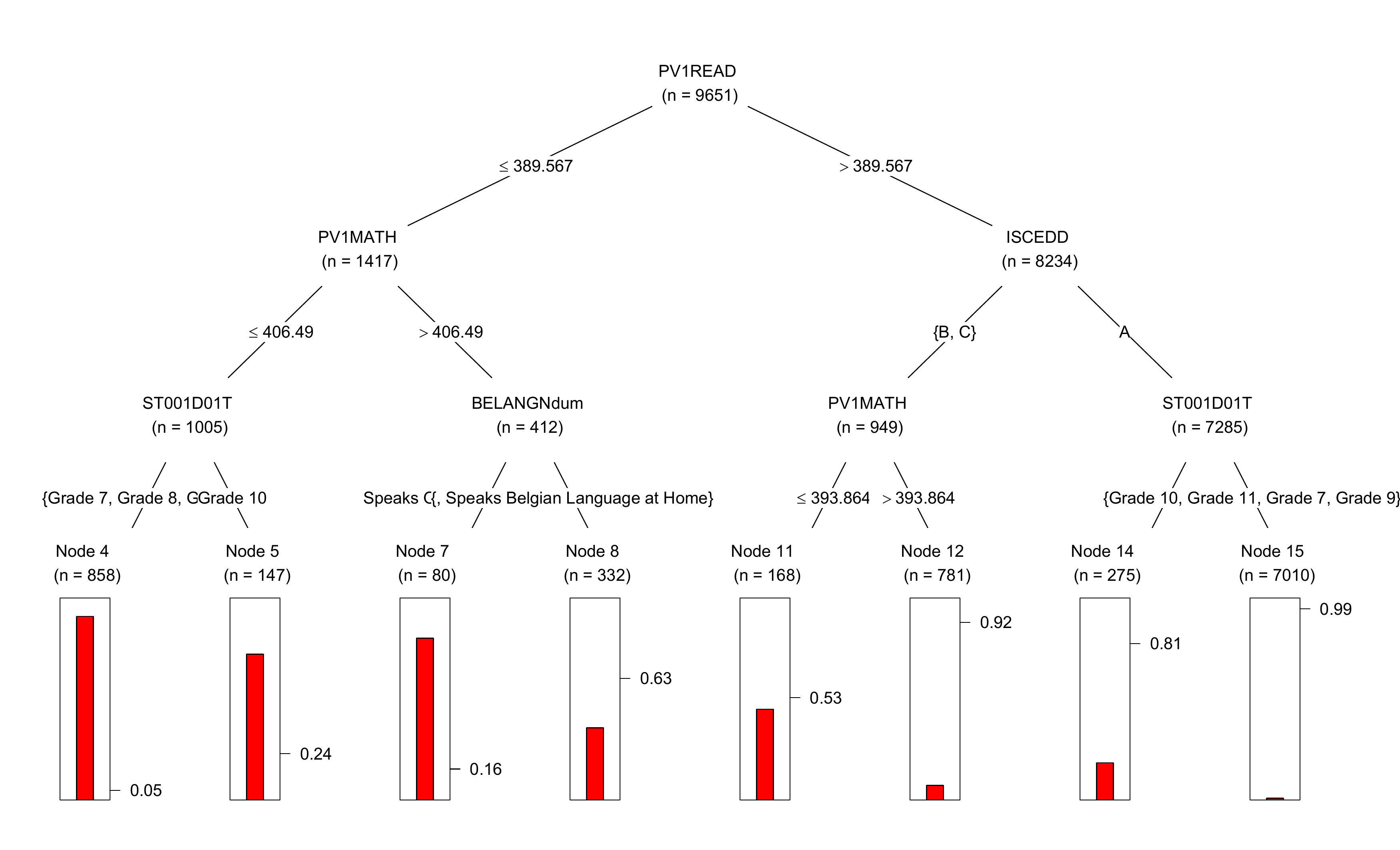}
    \caption{Conditional tree for the entire sample. Within each leaf is depicted in red the histogram of the percentage of units that have a low financial literacy score, and next to it the percentage of units with not-low financial literacy score within the same leaf.}
    \label{fig:ctree_general}
\end{figure}

\begin{figure}[H]
    \centering
    \begin{subfigure}[b]{0.48\textwidth}
    \includegraphics[width=\textwidth]{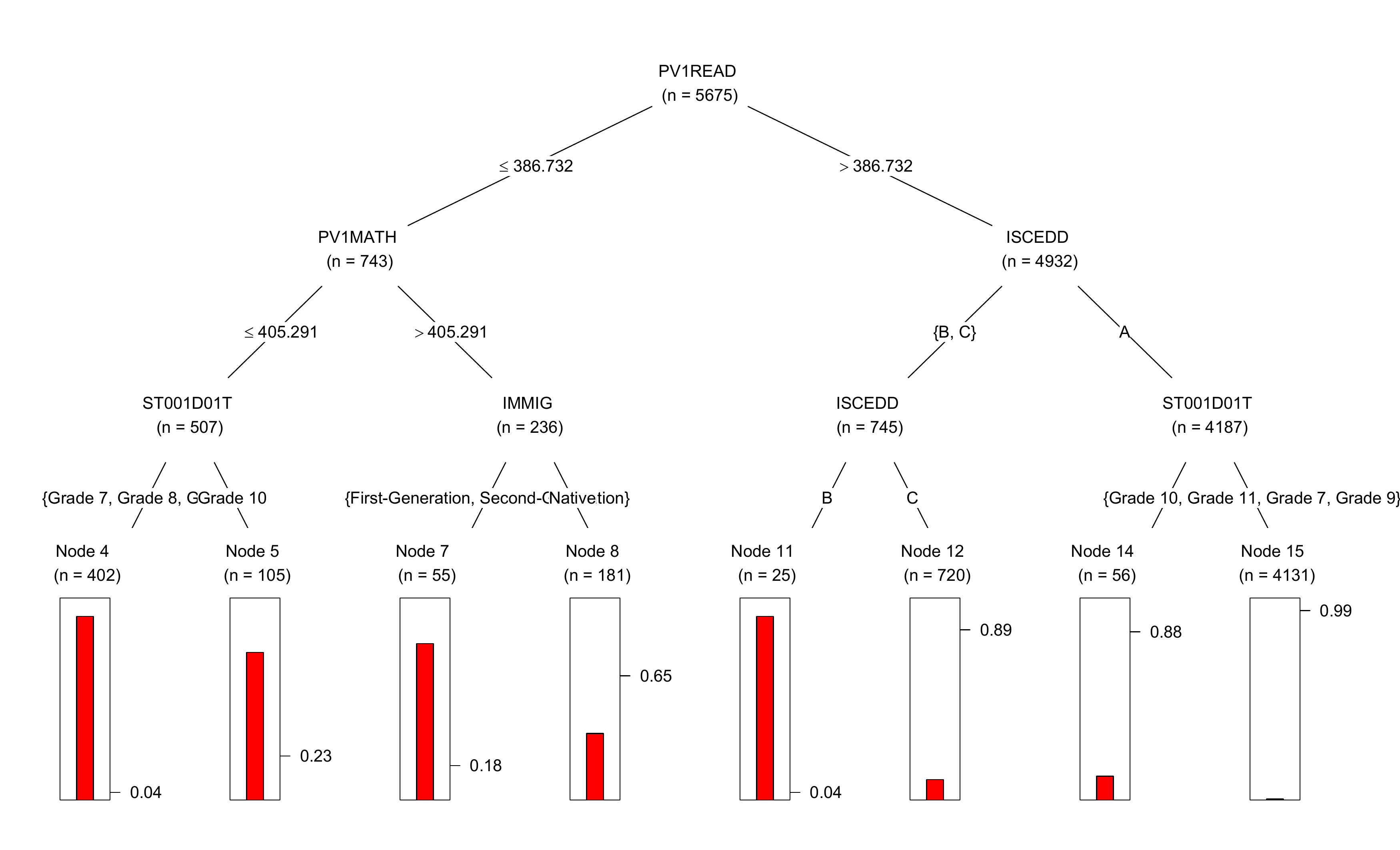}
    \end{subfigure}
    \begin{subfigure}[b]{0.48\textwidth}
    \includegraphics[width=\textwidth]{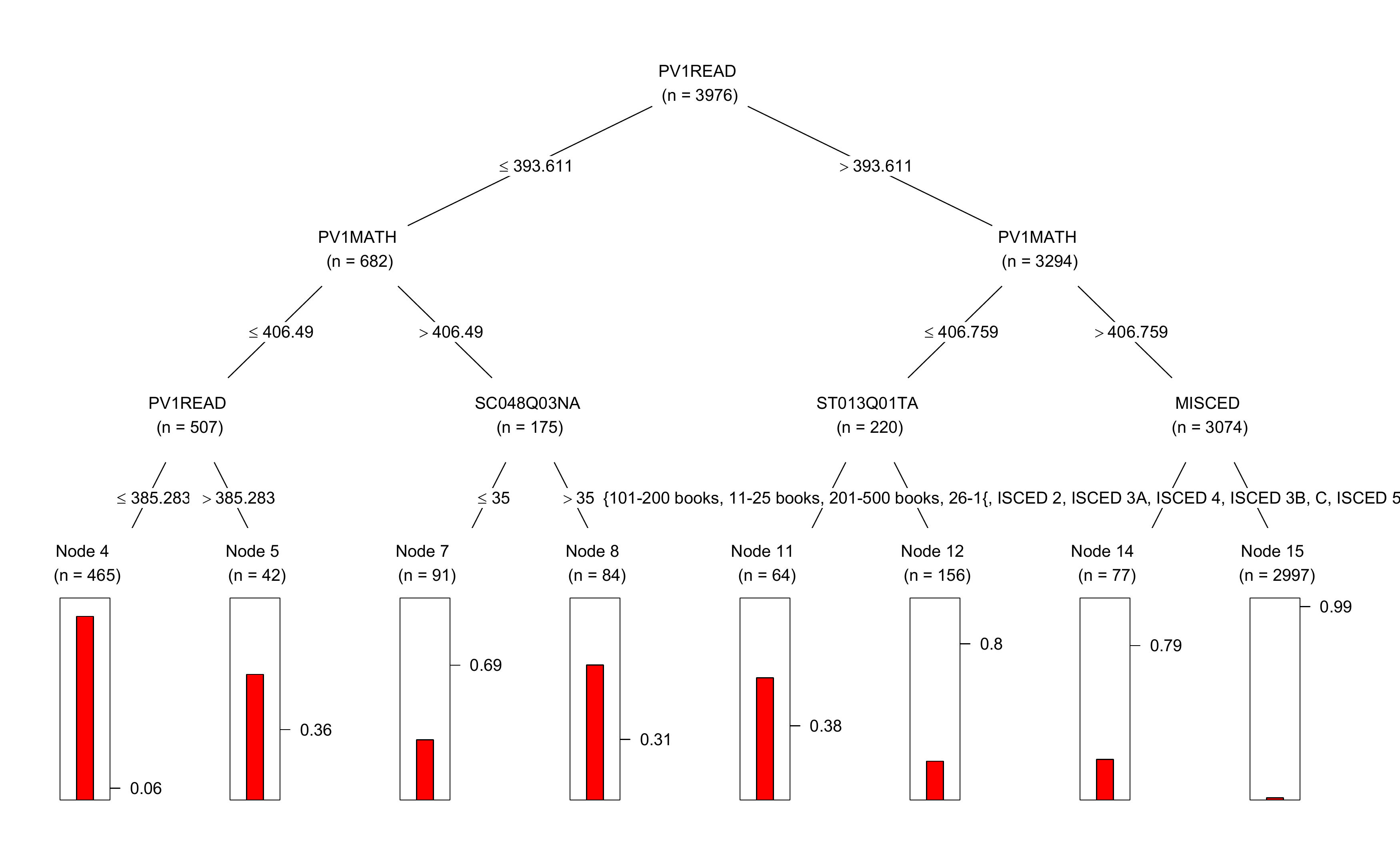}
    \end{subfigure}
    \caption{\footnotesize(Left) conditional tree for Flanders. (Right) The corresponding tree for Wallonia.}
    \label{fig:flanders_vs_wallonia1}
\end{figure}

\begin{figure}[H]
    \centering
    \begin{subfigure}[b]{0.48\textwidth}
    \includegraphics[width=\textwidth]{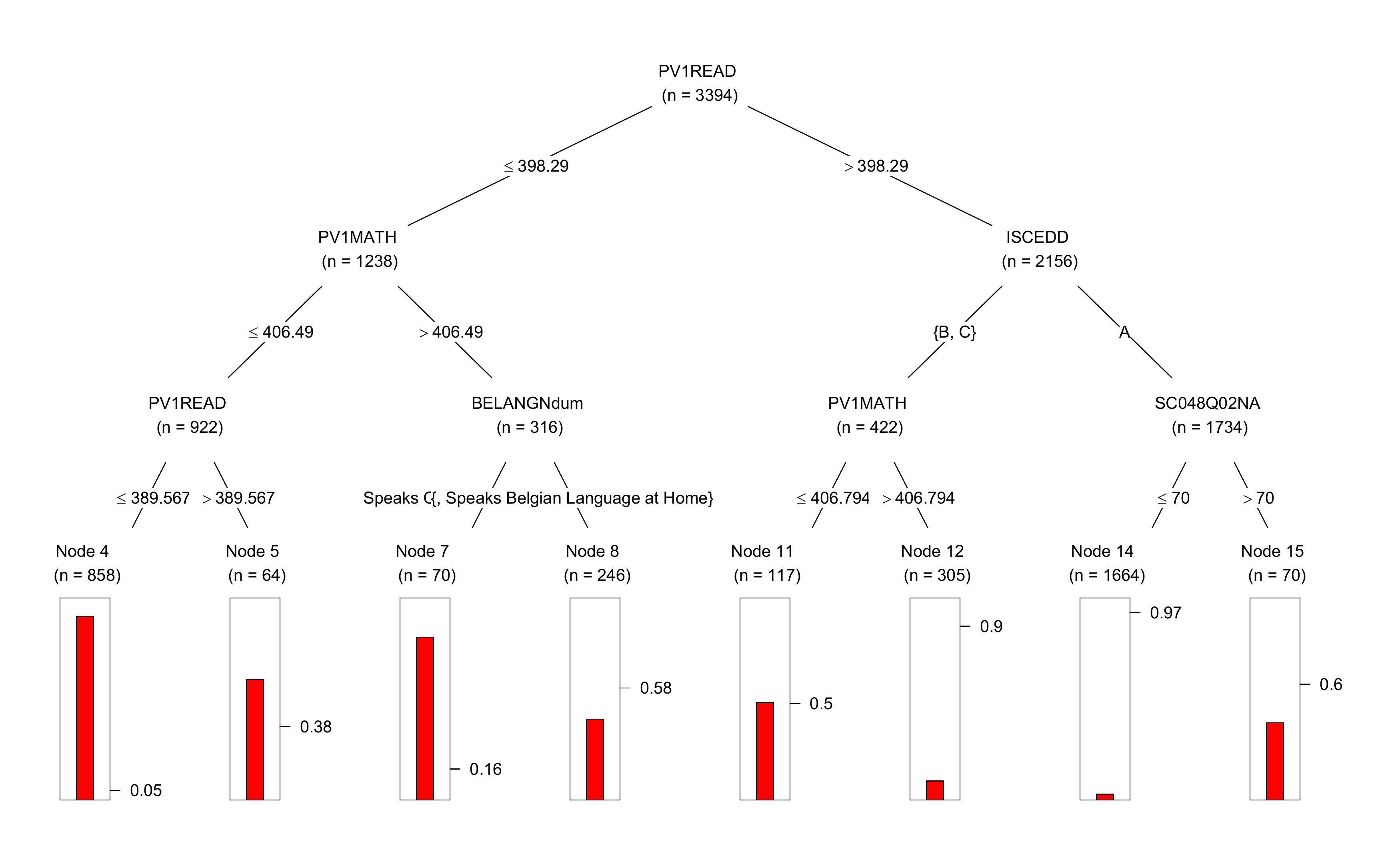}
    \end{subfigure}
    \begin{subfigure}[b]{0.48\textwidth}
    \includegraphics[width=\textwidth]{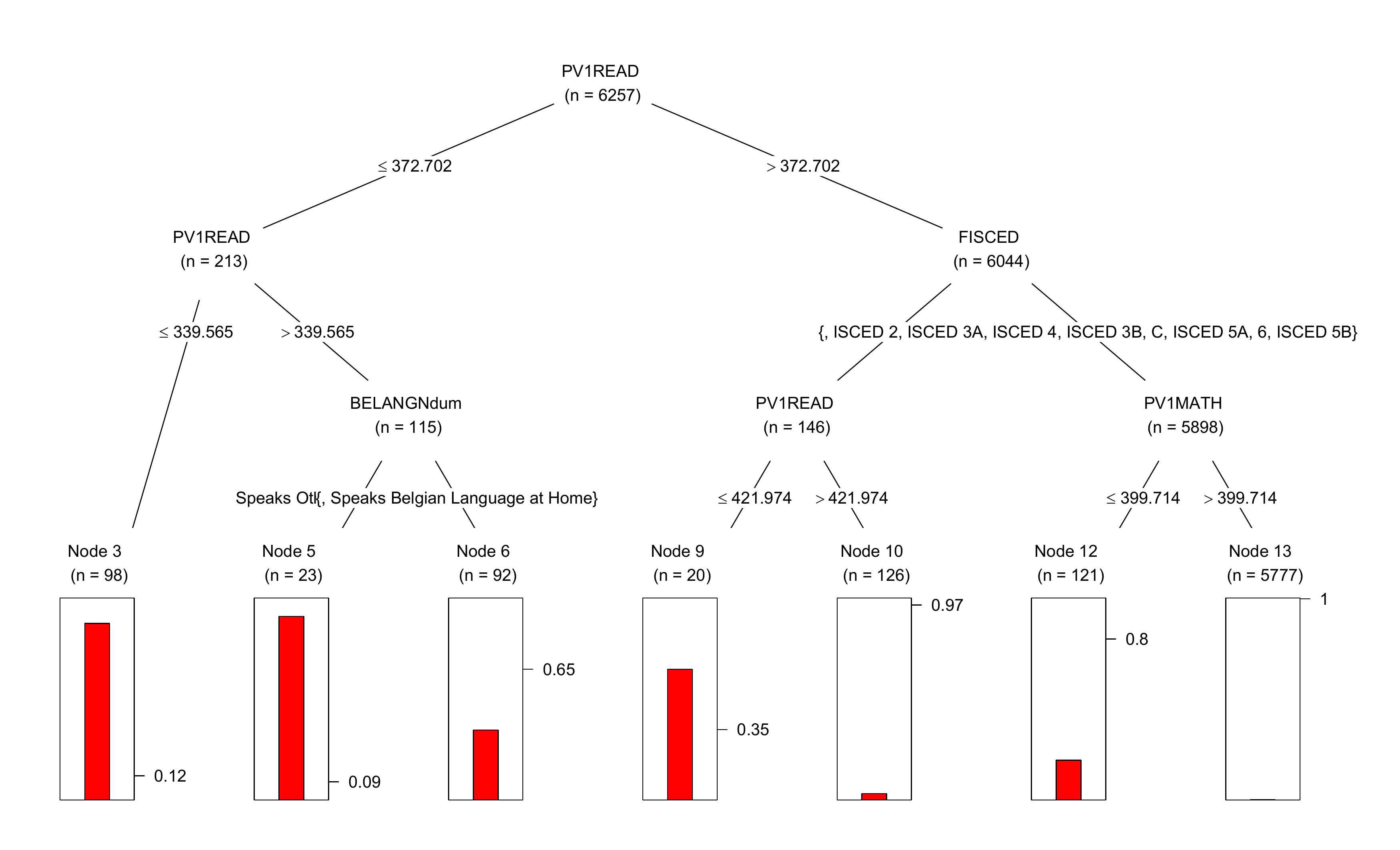}
    \end{subfigure}
    \caption{{\footnotesize(Left) conditional tree for students in grades 7 to 9. (Right) The corresponding tree for students in grades 10 to 12.}}
    \label{fig:ctree_grades1}
\end{figure}

\begin{figure}[H]
    \centering
    \begin{subfigure}[b]{0.48\textwidth}
    \includegraphics[width=\textwidth]{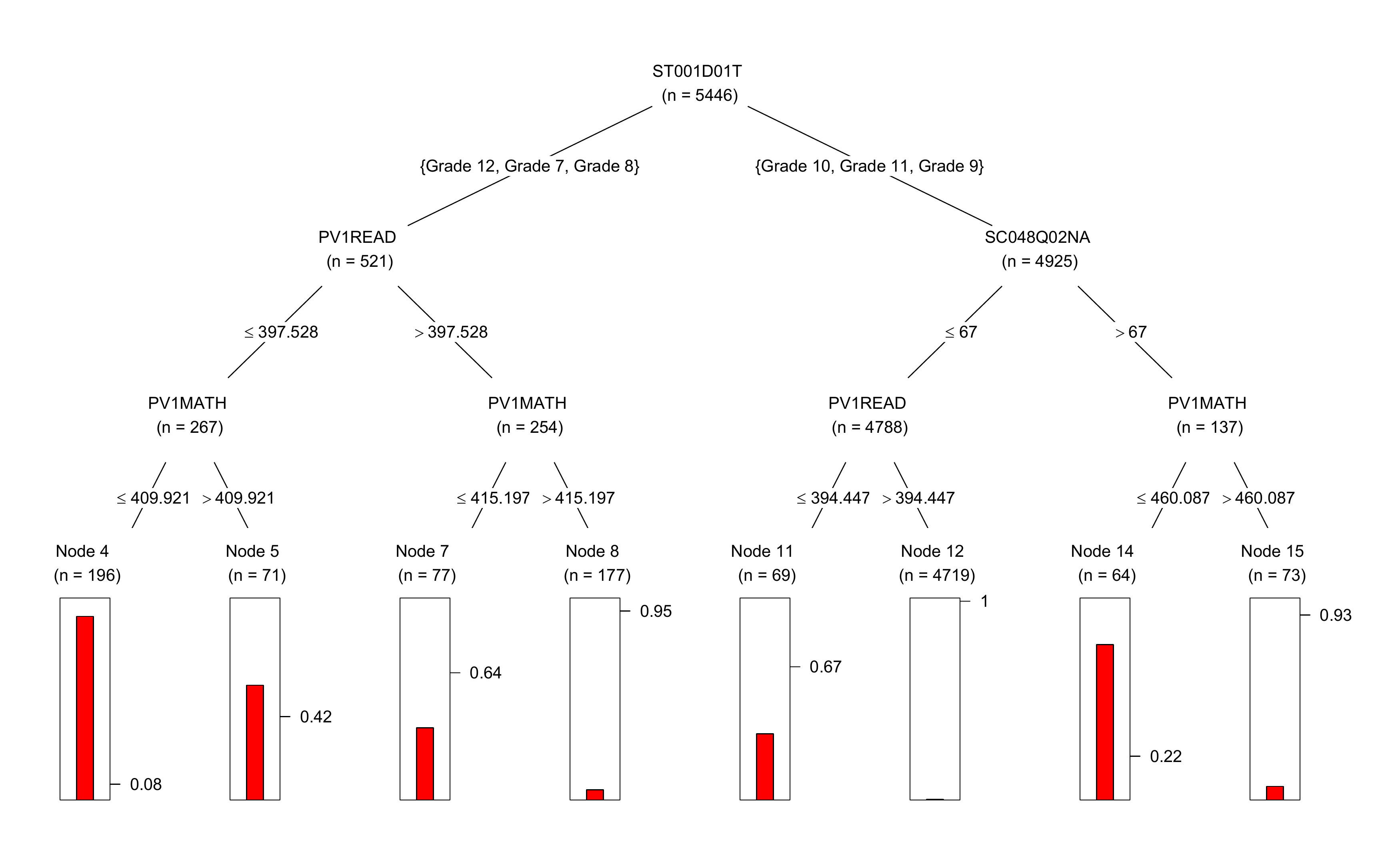}
    \end{subfigure}
    \begin{subfigure}[b]{0.48\textwidth}
    \includegraphics[width=\textwidth]{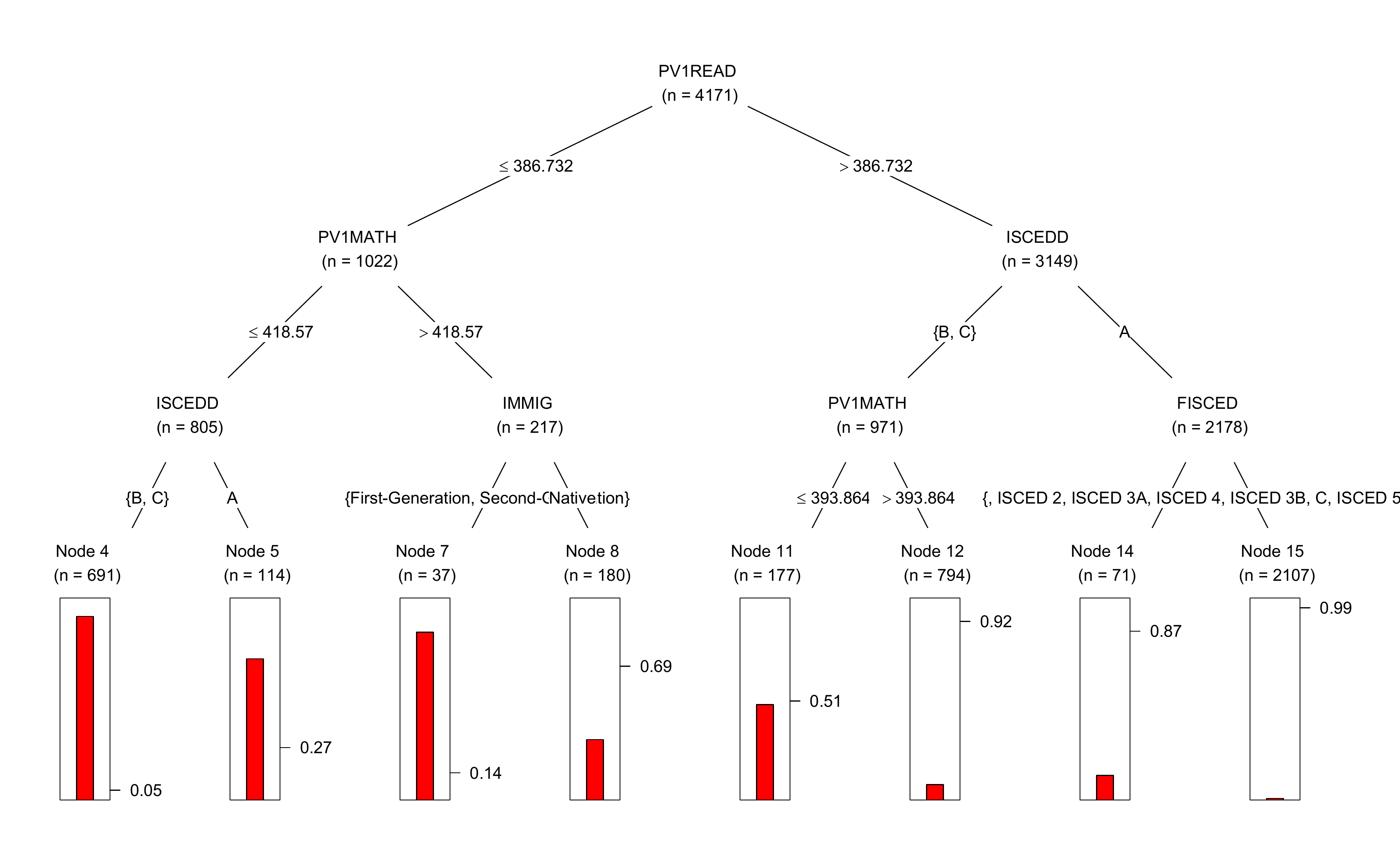}
    \end{subfigure}
    \caption{{\footnotesize(Left) conditional tree for general education. (Right) The corresponding tree for vocational education.}}
    \label{fig:ctree_vocational1}
\end{figure}

Figure \ref{fig:ctree_general} depicts the conditional tree for the overall sample, Figure \ref{fig:flanders_vs_wallonia1} shows the conditional trees for Flanders and Wallonia respectively, Figure \ref{fig:ctree_grades1} depicts the trees for students in grades 7 to 9 and 10 to 12, and Figure \ref{fig:ctree_vocational1} shows the results for the trees for students in general and vocational education.

As shown in Figure \ref{fig:ctree_general}, reading (\texttt{PV1READ}) and math  (\texttt{PV1MATH}) scores, the language spoken at home (\texttt{BELANGN}), study track (\texttt{ISCEDD}) and grade (\texttt{ST001D01T}) are important variables distinguishing groups of students in our entire sample. For students from the lower grades (grade 7 to 9) with lower scores on reading ($\leq389.567$) and math ($\leq406.49$) the predicted probability of low FLS is 95 percent. A higher PISA math  score ($>406.49$), but not being a native speaker results in a predicted probability of 84 percent of having low FLS. On the opposite side, students with a reading score above baseline ($>389.567$), following general programs designed to give access to the next level, have - depending on their grade level - a predicted probability of respectively only 19 or 1 percent of having low FLS.  

In Figure \ref{fig:flanders_vs_wallonia1}, we split by region. In the tree in the left panel of the Figure, we note that in Flanders, the largest predictive proportion of students with low FLS are again situated among those with lower scores for reading ($\leq386.732$) and math ($\leq405.291$). In the group with a low reading score ($\leq386.732$) and a higher math  score ($>405.291$) immigration status seems to matter (\texttt{IMMIG}). Among those who are first generation-migrants the predictive probability of having low FLS is 82 percent. In Wallonia, the proportion of students with low FLS is again the largest (94 percent) among those with low reading ($\leq385.283$) and math  scores ($\leq406.49$). In Wallonia (tree in the right panel), mother's education (\texttt{MISCED}) seems to play a distinctive role. Students with better reading and math  scores and with a highly educated mother only have a predictive probability of 1 percent of low FLS.

In Figure \ref{fig:ctree_grades1}, we split by grade. We group grades 7-9 (left panel), and 10-12 (right panel). Indeed in Flanders, after grade 9 certain conditions change (teacher qualification requirements, etc.). Hence, it is interesting to investigated the drivers of heterogeneous effects by grade. In grades 7-9, the largest predictive proportion (95 percent) of students with low FLS are again among those with a lower reading ($\leq398.29$) and math  score ($\leq389.567$). The second largest proportion of students with low FLS (84 percent) is found among non-native speakers with a high math  score ($\>406.49$) but with a low reading score ($\leq398.29$). The same pattern seems to exists in grades 10 to 12. However, in grades 10 to 12, the education of a student's father (\texttt{FISCED}) seems to have a significant influence on FLS. For students with fathers with a higher education background, the predictive probability of low FLS lies between 1 and 20 percent, depending on whether the student has a math  score above or below 399.714.

In Figure \ref{fig:ctree_vocational1}, we split by track. In the case of general education, students (left panel) from grade 7, 8 and 12 with a low reading ($\leq397.528$) and math  score ($\leq409.921$) have, with a predictive probability of 92 percent, the highest chance of having low FLS. The second largest predictive proportion (78 percent) is among students in grade 9, 10 and 11 in schools with more than 67 percent of special needs students (SC0848Q02NA) who have a math  score smaller or equal to 460.087. Considering only the students in vocational education (right panel), we observe that the main determinants of students' low financial literacy are the study track, math and reading scores and whether or not the student speaks Dutch at home.

\subsection{Results for sensitivity of predictions analysis}

In this Section, we report the results for the sensitivity of predictions analysis. In Section \ref{subsec:sensitivity} we introduced the methodology which we employ here to assess how an unobserved predictor, uncorrelated to the observed ones, could impact the predicted FLS and the performance of the model. In particular, the \texttt{sensitivity} function that we developed in the statistical software \texttt{R} identifies how many unit level predictions are statistically different between the original model and the model with the synthetic predictor and how much the synthetic predictor would improve the overall performance of the model.

Starting from the unit level predictions that we introduced in Section \ref{subsec:sensitivity}, we can get a sense of how much the synthetic predictor would affect the model prediction in a varying range of correlations between the synthetic predictor and the outcome (in the case of our application this range is  $[0.1, 0.5]$).\footnote{This range depends on the data at hand as we need to guarantee that $\mathbf{R}$ is a symmetric, positive definite matrix.} Here, we test how many of the predicted values of the augmented model would be statistically different from the predicted values of the original model. We find no evidence of statistically significantly different predictions up to a correlation of 0.5, where just 0.01\% of the unit level predictions of the augmented model are significantly different.

\begin{figure}[H]
    \centering
    \begin{subfigure}[b]{0.48\textwidth}
    \includegraphics[width=\textwidth]{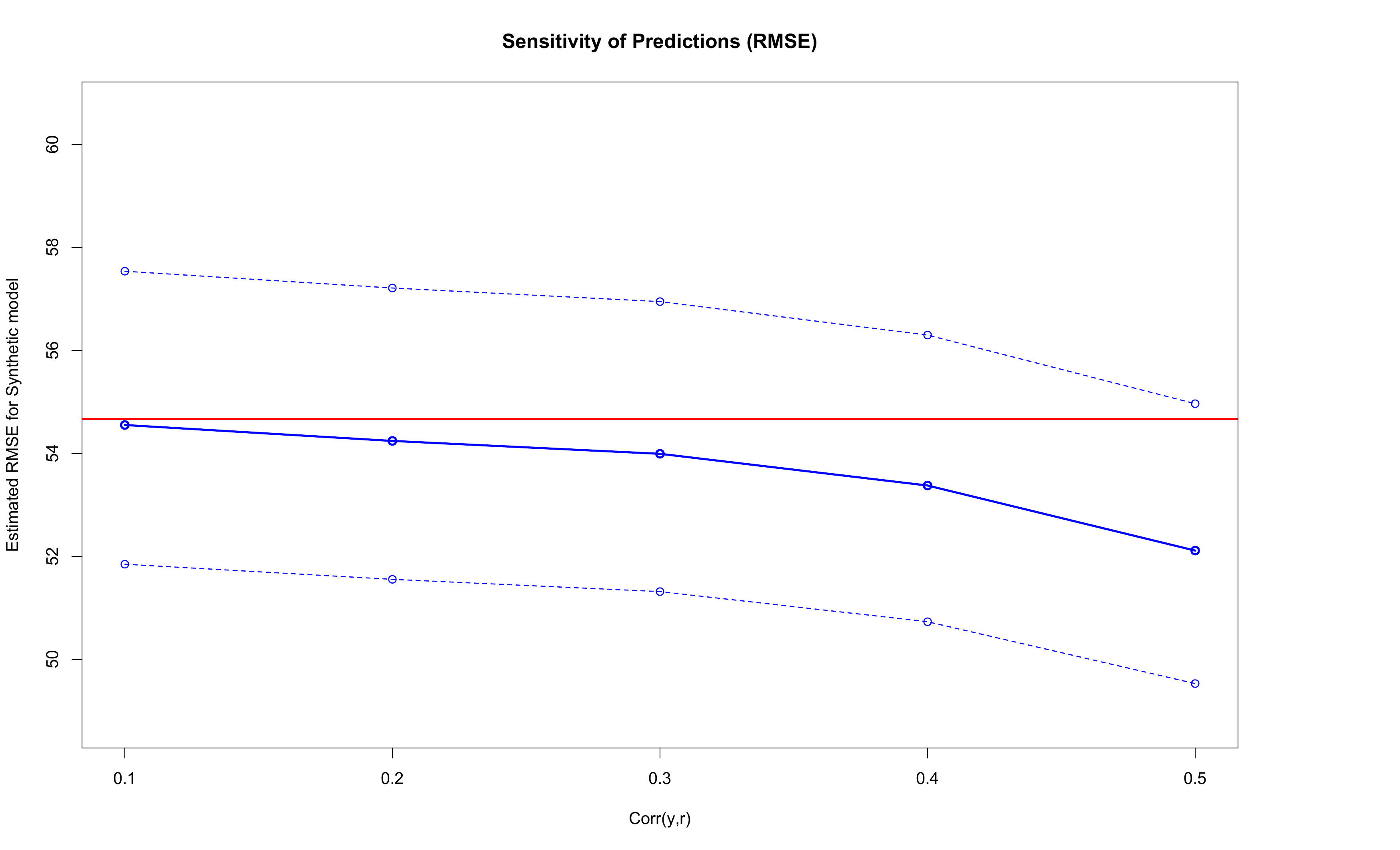}
    \end{subfigure}
    \begin{subfigure}[b]{0.48\textwidth}
    \includegraphics[width=\textwidth]{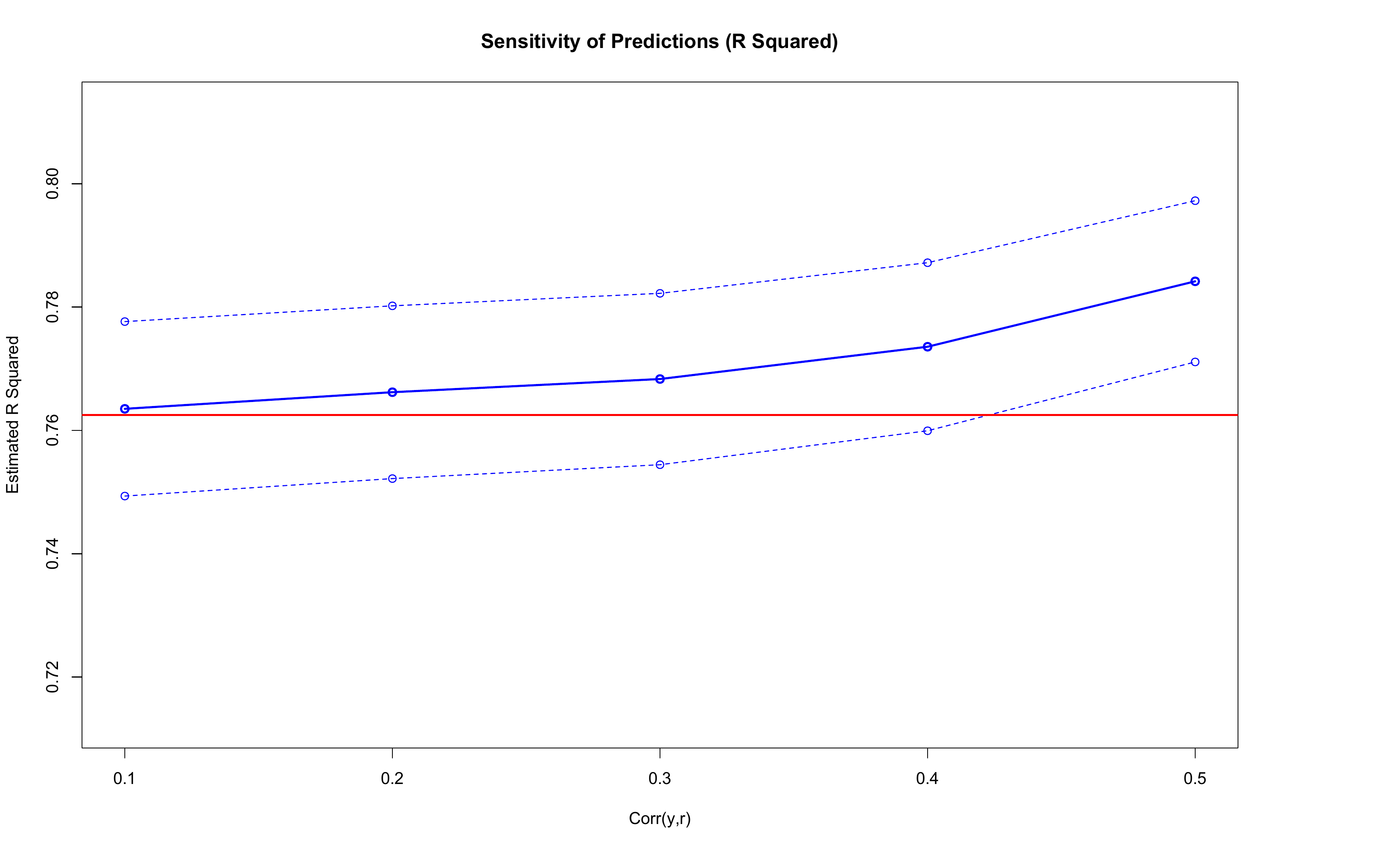}
    \end{subfigure}
    \caption{{\footnotesize(Left) Estimated RMSE for the augmented model with confidence intervals (blue) and for the original model (red). (Right) Estimated $R^2$ for the augmented model with confidence intervals (blue) and for the original model (red). The 99\% confidence intervals were constructed following \cite{cohen2014applied}.}}
    \label{fig:sensitivity}
\end{figure}

Moreover, figure \ref{fig:sensitivity} depicts the results for both the RMSE and $R^2$ of the original and augmented model with their confidence intervals as the correlation between the synthetic predictor $R$ and the outcome $Y$ increases. In both cases, even when the synthetic predictor has a correlation of 0.5 with the outcome, there is no significant difference between the RMSE of the original and the augmented model. The $R^2$ is significantly improved just when the correlation reaches 0.5. This evidence could be interpreted as the fact that, in order to get a significant increase in the $R^2$, we would need an explanatory variable (completely unrelated to the observed ones) with at least a 0.5 correlation to the outcome. To put things into perspective, a predictor with such a correlation would rank between the first three best predictors in the model. We argue that, in the context of our application, such an unobserved variable would be very hard, if not impossible, to find.  

As argued before, the interpretation of the sensitivity of prediction analysis is that, even if we could introduce a new, unobserved predictor, with a high explanatory power in the model, the unit level predictions as well as the overall performance of the model would remain fairly stable. This set of evidence, together with the non-significant variation in the unit level predictions, hints at the fact that our model is already capturing most of the signal in the data and its predictions are stable. Hence, this analysis offers to policy-makers an additional evidence in favour of using the predictions from the model that we propose to enact targeted policies.

Moreover, we make this analysis robust by introducing correlations between the synthetic predictor and the two best predictors.\footnotetext{The best predictors are chosen according to the built-in variables' importance measure of BART. For a review of various tree-based variables' importance measures and their pitfalls we refer the reader to \cite{james2013tree} and the recent contribution by \cite{gottard2020note}.} The size of the correlation is the same as the correlation introduced between the synthetic predictor and the outcome. In the case of our application, the model is robust because the results from the sensitivity analyses are invariant to the introduction of this additional correlation patterns.

\section{Discussion and conclusions} \label{sec:discussion}

In this paper, we introduce a novel sensitivity analysis to assess the robustness of our preferred predictive model in terms of the stability of its predictions and performance. The model that we propose is general enough to be applied to both Bayesian techniques (such as the Bayesian Additive Regression Trees algorithm) or frequentist techniques (such as the Random Forest algorithm). In particular, we propose a novel way to partially answer policy-relevant questions such as: ``\textit{how much would an unobserved predictor impact the model's prediction and its performance?}". To do so, we develop a novel methodology to construct a synthetic predictor that is correlated with the outcome but is uncorrelated with the observed predictors. By generating the synthetic predictor in this a way, we are able to interpret the results from the sensitivity analysis in a way that hints at how much an exogenous predictor with a high explanatory value would add to the model. If the addition of the synthetic predictor is not resulting in significant differences, then we can safely assume that our predictive model is capturing enough signal from the training data. Moreover, through our novel methodology, we are able to use a wide range of correlations between the outcome and the synthetic predictor. This enables the researchers to check whether or not there is a \textit{break-point} (given by the explanatory power of the synthetic predictor) above which the performance of the model would be significantly better. 

We apply our approach to predict FLS for students in a region of Belgium where FLS are unobserved. Next, we identify the characteristics of students with predicted FLS lower than the mean. Finally, we verify the sensitivity of our observations for unobserved predictors.
The first stage of our analysis provides evidence that the Bayesian Additive Regression Tree (BART) approach outperforms the traditional random forest (RF) approach on a number of selected performance measures (i.e. smaller RMSE and MAE, higher $R^2$)   with respect to FLS predictions. We find that the predicted values for FLS for students in Flanders are, on average, somewhat larger than those for students in Wallonia. Nevertheless, the distribution of FLS in Flanders is more extreme (i.e. lower minimum, larger maximum) compared to Wallonia. 

Next, we estimate the posterior predicted observations available for each student which also allows to identify outliers (i.e. students for which the the FLS are situated outside the 95\% credibility intervals for the mean values). Our results show that the predictive probability of having low FLS is the largest for students with lower scores on reading and math in the PISA test. This corroborates with findings of \cite{Mancebon2019} who argue that the development of financial abilities of students is mediated by their mathematical skills. Another interesting observation is that the family background also has a key influence. Students with the largest predicted low FLS are often students from families where the school language is not spoken at home. Particularly in Flanders, being a first-generation immigrant or not is key in having low FLS. The educational background of parents is another important predictor. These observations are in line with existing literature suggesting that more vulnerable groups in terms of low FLS are often situated among those with a lower socioeconomic background \citep{Gramatki2017, Riitsalu2016,DeBeckker2019}. 

Finally, we confirm that our novel approach is robust in terms of potential unobserved variables. The inclusion of a synthetic predictor highly correlated with the outcome does result in a significant difference in the performance measures of the original model and the augmented model.
From a policy perspective, our novel approach is highly interesting as it not only allows to predict outcomes for certain countries (or regions within countries, as in the case of our application) with missing data, but also introduces the possibility to detect observations that fall outside the mean outcomes. In this respect, policy-makers can detect which factors drive low results. We applied our model to PISA financial literacy data, however, the same approach can also be applied to other large administrative datasets. 

\singlespacing
\bibliographystyle{dcu}
\bibliography{biblio}

\newpage
\appendix

 \pagenumbering{arabic}
    \setcounter{page}{1}
\setcounter{equation}{0}
\setcounter{table}{0}
\renewcommand{\thetable}{A\arabic{table}}
\setcounter{figure}{0}
\renewcommand{\thefigure}{A\arabic{figure}}
\begin{center}
\text{\LARGE \bf Online appendix}
\end{center}
\vspace{-0.9cm}\section{Data}\label{subsec:app-data}
\singlespacing
\begin{table}[H]
\centering
\caption{Variables used from the PISA data}
\label{tab:variables}
\begin{tabular}{@{}ll@{}}
\toprule
PISA Code &  Variable \\ 
\midrule
\textit{Student Characteristics} &   \\ 
\texttt{ST001D01T}         & International Grade                     \\
\texttt{ST004D01T}         & Gender           \\
\texttt{AGE}               & Age  \\
\texttt{ISCEDD}            & Study Track: ISCED Designation                   \\
\texttt{ISCEDO}            & Study Track: ISCED Orientation                    \\
\vspace{0.2cm}
\texttt{BELANGN}           & Speaks Belgian Language at Home    \\
\textit{Socioeconomic Status} &    \\  
\texttt{HEDRES}            & Educational Resources at Home    \\
\texttt{WEALTH}            & Family Wealth Index (Economic Possessions)        \\
\texttt{ST013Q01TA}        & Number of Books at Home  \\
\texttt{IMMIG}             & Immigration Status               \\
\texttt{MISCED}            & Mother's Education (ISCED)             \\
\texttt{FISCED}            & Father's Education (ISCED)           \\
\texttt{BMMJ1}             & Mother's Job (ISEI)                 \\
\texttt{BFMJ2}             & Father's Job (ISEI)          \\
\vspace{0.2cm}
\texttt{EMOSUPS}           & Parents Emotional Support      \\  
\textit{Achievement and Attitude} &     \\  
\texttt{PV1MATH}           & Plausible Value 1 in Mathematics        \\
\texttt{PV1READ}           & Plausible Value 1 in Reading           \\
\texttt{REPEAT}            & Grade Repetition  \\
\texttt{OUTHOURS}          & Out-of-School Study Time per Week  \\
\texttt{MMINS}             & Mathematics Learning Time at School           \\
\texttt{LMINS}             & Language Learning Time at School              \\
\texttt{ANXTEST}           & Personality: Test Anxiety      \\
\vspace{0.2cm}
\texttt{MOTIVAT}           & Achievement Motivation            \\   
\textit{School Characteristics} &    \\  
\texttt{SC001Q01TA}        & School Community (Location)    \\
\texttt{SC048Q01NA}   & Share of Students With a Different Heritage Language \\
\texttt{SC048Q02NA}        & Share of Students With Special Needs          \\ 
\texttt{SC048Q03NA} & Share of Socioeconomically Disadvantaged Students  \\
\texttt{SCHSIZE}           & School Size             \\
\texttt{CLSIZE}            & Class Size                           \\
\texttt{RATCMP1}           & Number of Available Computers per Student      \\
\texttt{LEADPD}            & Teacher Professional Development           \\
\texttt{SCHAUT}            & School Autonomy                 \\
\texttt{EDUSHORT}          & Shortage of Educational Material       \\
\texttt{STRATIO}           & Student-Teacher Ratio                   \\
\addlinespace  \bottomrule
\end{tabular}
\end{table}

\begin{center}
\input{Tables/summarystats.tex}\label{subsec:sumstats}
\end{center}

\vspace{3cm}

\begin{center}
\input{Tables/outcome.tex}
\end{center}

\vspace{3cm}

 \begin{figure}[H]
    \centering
    \resizebox{1.1\textwidth}{!}{
    \includegraphics{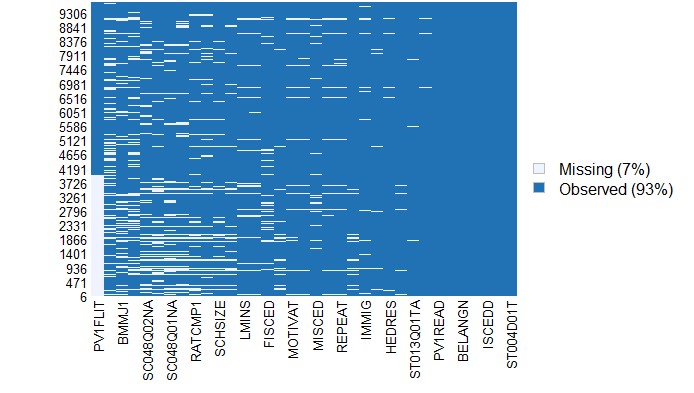}
    }
    \caption{Missingness Map}
    \label{fig:missingness}
\end{figure}

\begin{figure}[H]
    \centering
    \resizebox{0.9\textwidth}{!}{
    \includegraphics{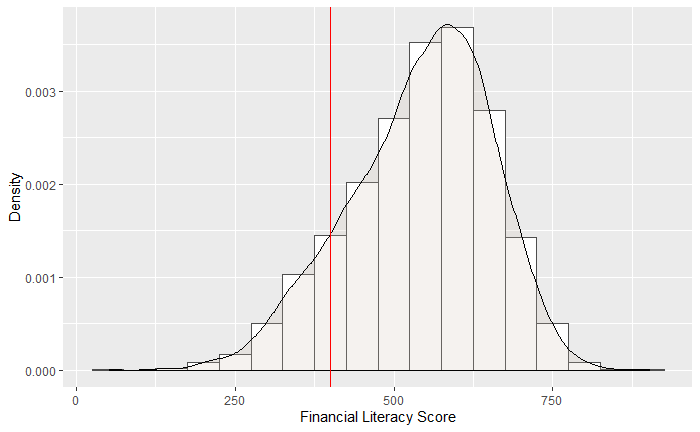}
    }
    \caption{Histogram of the Outcome Variable (PISA Financial Literacy Score). The red line indicates the threshold of the baseline level of proficiency in financial literacy. The OECD suggests that students above this threshold of 400 points have financial literacy levels that are sufficient to participate in society \citep{OECD2017a}.}
    \label{fig:outcome}
\end{figure}

\begin{figure}[H]
    \centering
    \resizebox{0.9\textwidth}{!}{
    \includegraphics{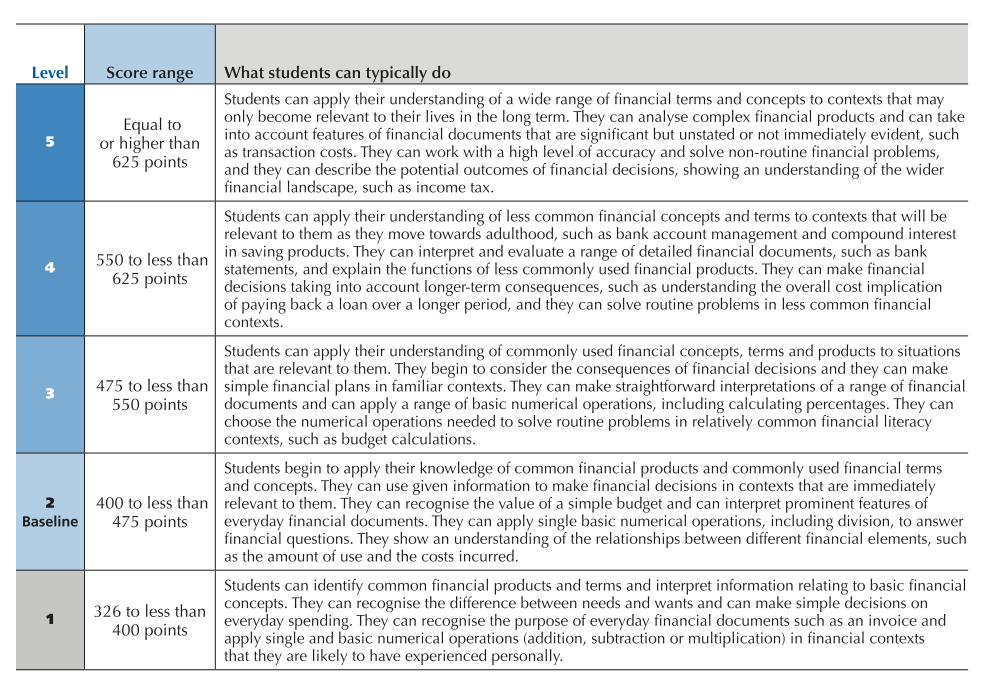}
    }
    \caption{PISA Proficiency Levels for Financial Literacy \citep{OECD2017a}}
    \label{fig:PISAlevels}
\end{figure}

\doublespacing

\section{Overlap} \label{subsec:overlap}

Researchers often face potential pitfalls of machine learning predictions when the data on which they train their algorithm are substantially different from the data on which they perform the predictions. We provide a novel methodology to investigate if there are regions of the predictors' space in the out-of-sample set where the algorithm is relying on extrapolation.

In real world scenarios, one may be facing the problem of generalizing the results from one sample -- for which one observes the distribution of a certain outcome -- to a novel sample for which one has no information on the outcome. Generalizing the prediction from the first sample (in-sample predictions) to the new sample for which the outcome is not observed (out-of-sample predictions) can potentially lead to less reliable predictions if there is scarce overlap in the observed predictors in the two samples. Lack of overlap in the support of in-sample and out-of-sample predictors leads to the predictor model heavily relying on extrapolation in the non-overlap areas.

As highlighted by \cite{hooker2004diagnostics}, extrapolation can create unrealistic predictions, even in reasonable points of the predictors space. Hence, we argue that it is critical to get a clear sense of how much the predictive model at hand is relying on extrapolation. As a rule of thumb, we argue that one should rely less on predictions obtained in subsamples for which the overlap is small.

To evaluate the performance of an algorithm on a new sample different from the sample on which the algorithm is trained, commonly, a simple random sampling is performed which divides the {\em study dataset} $\Omega_{s}$ into two samples. One sample, the training sample $\Omega_{s,tr}$, is used for the construction of the algorithm. The other sample, the test sample $\Omega_{s,te}$, is used to assess the performance of the algorithm Performing random sampling guarantees that the observable predictors in $\Omega_{s,tr}$ and $\Omega_{s,te}$ are fairly balanced, i.e., the distributions for the predictor in $X_{\Omega_{tr}}$ and $X_{\Omega_{te}}$ are similar in both samples. 

In real world scenarios we may be facing the problem of generalizing the results from one sample for which we observe the distribution of a certain outcome ($\Omega_s$), to a {\em target sample} for which we have no information on the outcome ($\Omega_t$). However, the predictive performance of the algorithm may not be mimicked when the characteristics of the new sample $\Omega_t$ are fairly different from the characteristics of $\Omega_s$. For example, this can be the case if $\Omega_t$ and $\Omega_s$ are not random subsamples from the same super-population.
In order to tackle this problem we propose to rely on what we call \textit{overlap} score.

\subsection{Overlap Score}

We define the overlap score as the probability of being in the study dataset conditional on a certain set of predictors. In mathematical terms this probability is the following:
\begin{equation*}
o(\bX)=Pr(\mathbb{1}(S)=1|\bX=x),
\end{equation*}
where:
\begin{equation*}
    \mathbb{1}(S)= 
    \begin{cases} 
    1 & \text{if}  \:\:\: i \in \Omega_s; \\
    0 & \text{otherwise}.
    \end{cases}
\end{equation*}
$\hat{o}(\bX)$ can be estimated using generalized linear models for binary outcome such as logistic or probit regressions or machine learning methods for classification such as classification tree \citep{friedman1984classification} and random forest \citep{breiman2001random}.

We argue that the overlap score provides a useful insight to understand how much the machine learning technique is relying on extrapolation to extend the predictions from the study sample to the target sample. In principle, the idea would be to check if, for any observation in the study population, there is a similar counterpart in the target population. This may be done in various ways by defining different distance measures (e.g., Mahalanobis distance). The nice feature of the overlap score is that it furnishes an easy and computationally scalable way of performing such an analysis. Indeed, the overlap score can be thought of as a measure of the distance between two units in the study and in the target population. In this, the overlap score is similar to its causal inference counterpart, the propensity score \citep{rosenbaum1983central}.

\subsection{Results for multivariate overlap}

Here, we propose to use the overlap score to assess if there are regions of the features spaces where there is no overlap between students in Flanders and in Wallonia. This can be easily done by checking whether there are units in the study population with an estimated overlap score lower (greater) than the lowest (greatest) overlap score in the target population. The algorithm that we propose, is implemented in such a way that it gives the researcher the possibility to trim away these non-overlapping units.

\begin{figure}[H]
    \centering
    \includegraphics[width=\textwidth]{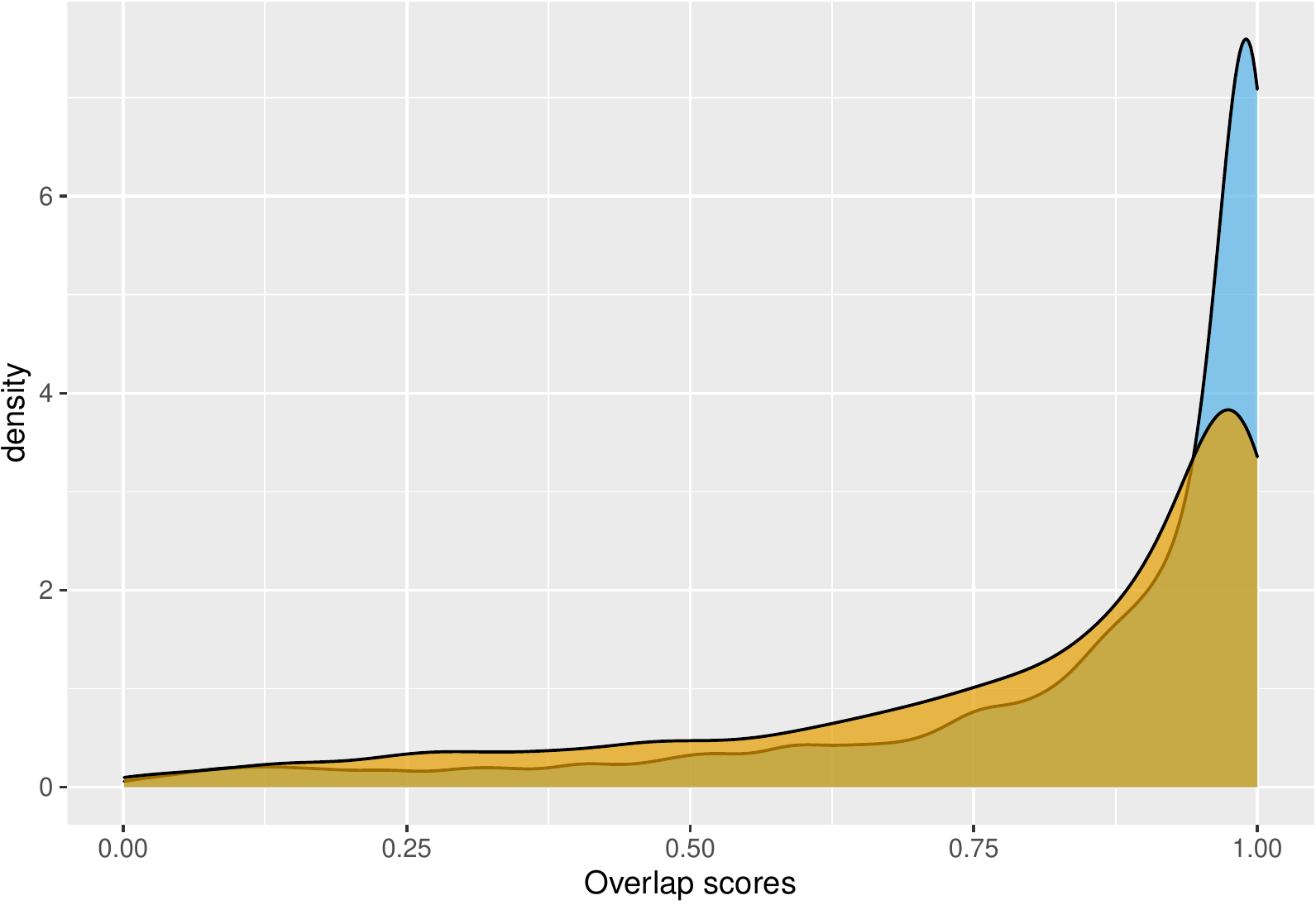}
    \caption{{Predicted overlap scores for Flanders (light blue) and Wallonia (orange).}}
    \label{fig:overlap}
\end{figure}

Figure \ref{fig:overlap} depicts the results for the overlap score obtained fitting a logistic regression. In light blue the predicted overlap scores for students in Flanders, while in orange the complement of predicted overlap scores for students in Wallonia. As one can see, there is wide overlap between the two distributions hinting at the fact that there are no regions of the features' space where the predictive algorithm will rely on extrapolation to perform out-of-sample predictions.

\section{Robustness check}\label{subsec:robustness}

To perform this robustness check, we use a reworked version of the 10-folds cross-validation used in the previous Section. This procedure assigns nine random folds from the Flemish data to the training sample and one random fold from the Walloon data to the testing sample at each split. The performance of the method is evaluated at each split and then the results are aggregated as a usual 10-fold cross-validation.

The results are depicted in Table \ref{table:performance_outcomes}. The adjusted $R^2$ for BARTs built on the different outcomes are comparable to the ones in Table \ref{table:performance} and hint at the fact that there are no relevant differences between the performance of the model built using the Flemish data both for training and testing and the one built using Flemish data for training and Walloon data for testing. In turn, this robustness check seems to confirm the fact that the `technology' and the efficiency in Flanders and Wallonia are comparable.

\vspace{1cm}
\begin{table}[H]
\centering
\small
\caption{Summary statistics of the performance of the BART on different outcomes}
\label{table:performance_outcomes}
\begin{tabular}{l*{1}{ccc}}
\toprule
                &     \textit{RMSE} & \textit{MAE} &  $R^2$ \\
\midrule
\textit{Math Score Outcome} &   52.5344 &   41.8943 &    0.7099 \\
\textit{Reading Score Outcome} &   52.5421 &   41.8146 &    0.7064 \\
\bottomrule
\end{tabular}
\end{table}

\section{Changing outliers definition} \label{appendix:outliers}
\doublespacing

Below we depict the results for the conditional tree analysis, when we define outliers as observation with predicted values below two absolute deviation from the median. This is a more restrictive definition of outliers. 

\begin{figure}[H]
    \centering
    \includegraphics[width=1\textwidth]{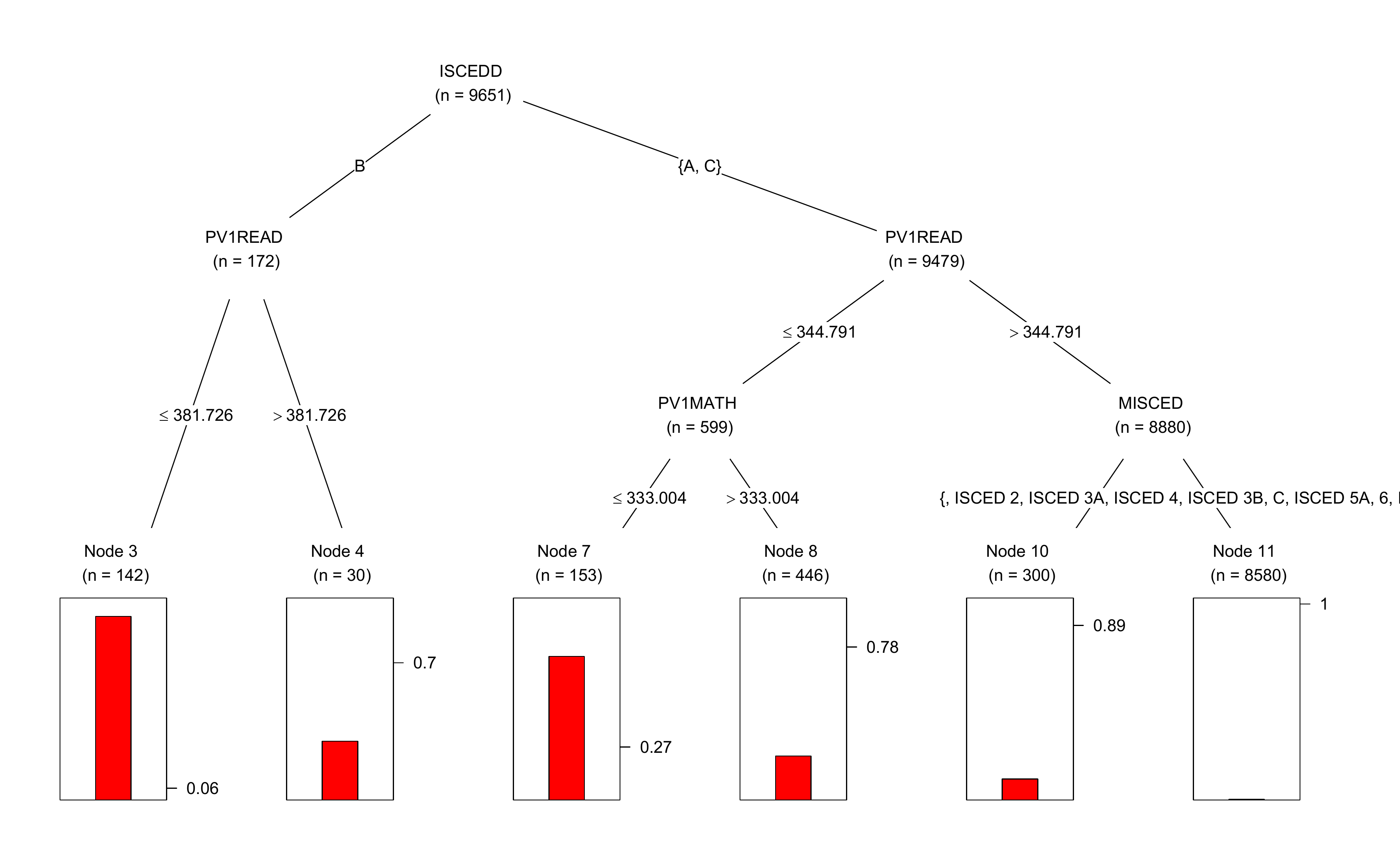}
    \caption{Conditional tree for the entire sample. Within each leaf are depicted in red the histogram of the percentage of units that have a low financial literacy score, and next to it the percentage of units with not-low financial literacy score within the same leaf.}
    \label{fig:ctree_general_mad}
\end{figure}

\begin{figure}[H]
    \centering
    \begin{subfigure}[b]{0.48\textwidth}
    \includegraphics[width=\textwidth]{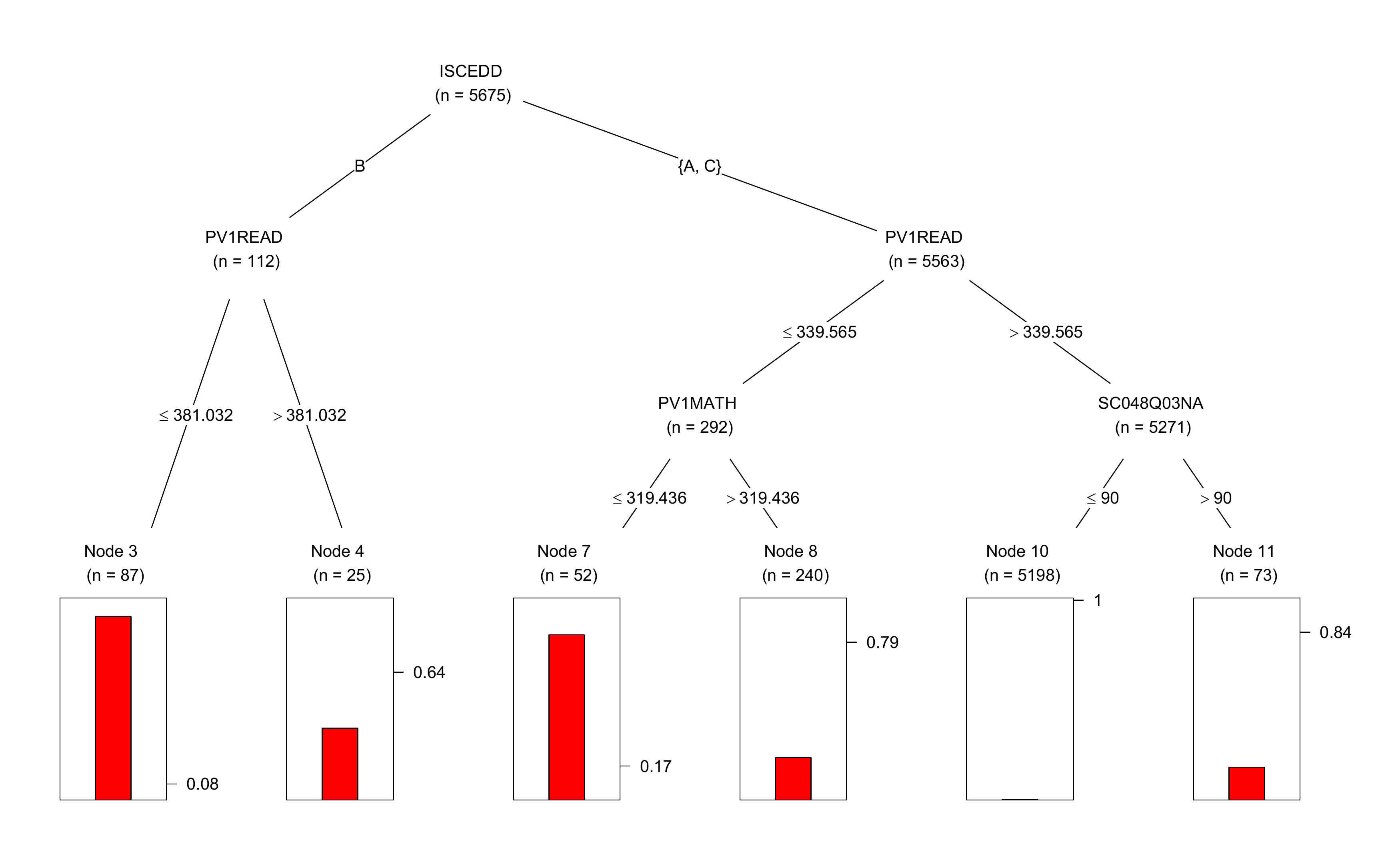}
    \end{subfigure}
    \begin{subfigure}[b]{0.48\textwidth}
    \includegraphics[width=\textwidth]{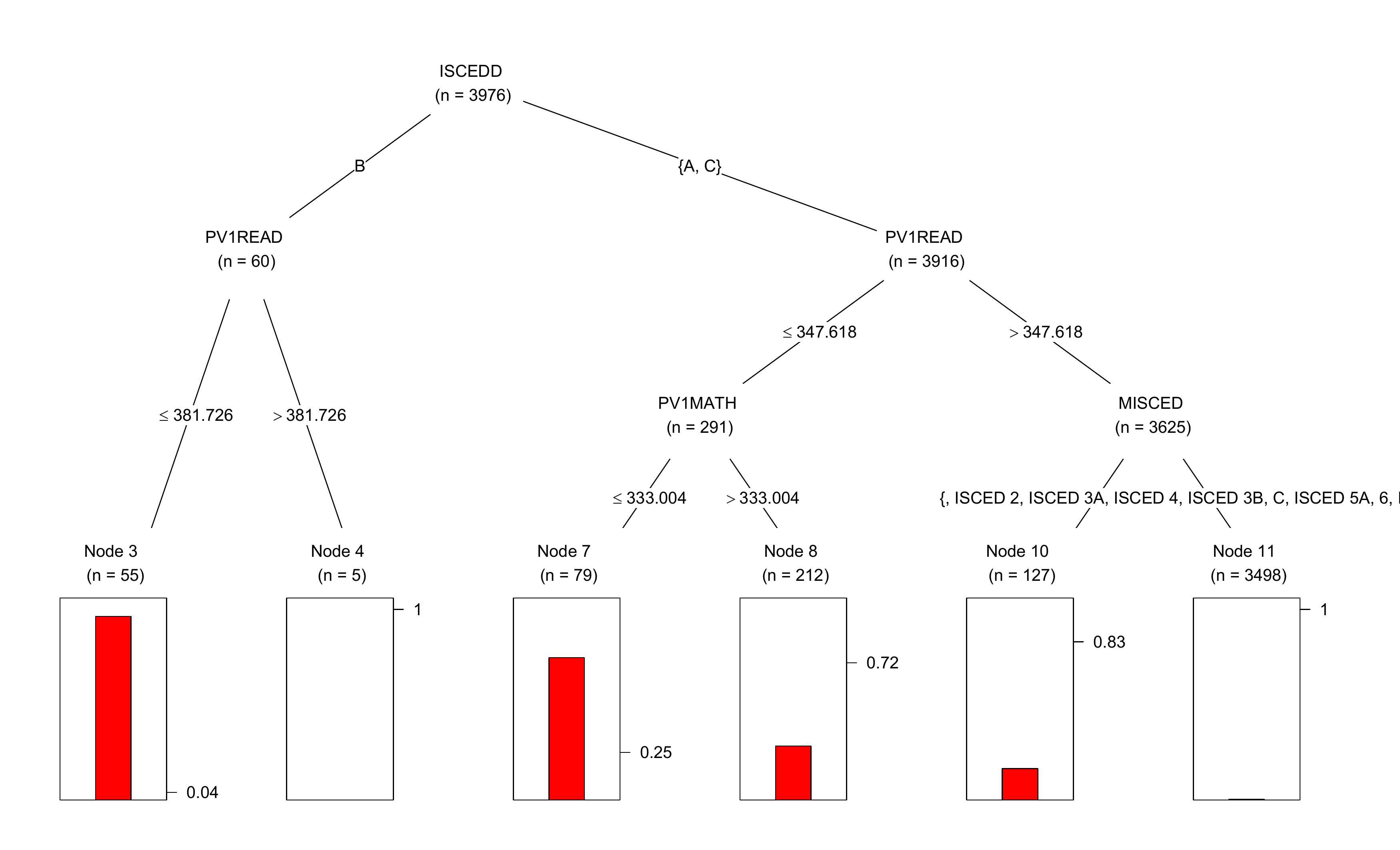}
    \end{subfigure}
    \caption{{\footnotesize(Left) conditional tree for Flanders. (Right) The corresponding tree for Wallonia.}}
    \label{fig:flanders_vs_wallonia}
\end{figure}

\begin{figure}[H]
    \centering
    \begin{subfigure}[b]{0.48\textwidth}
    \includegraphics[width=\textwidth]{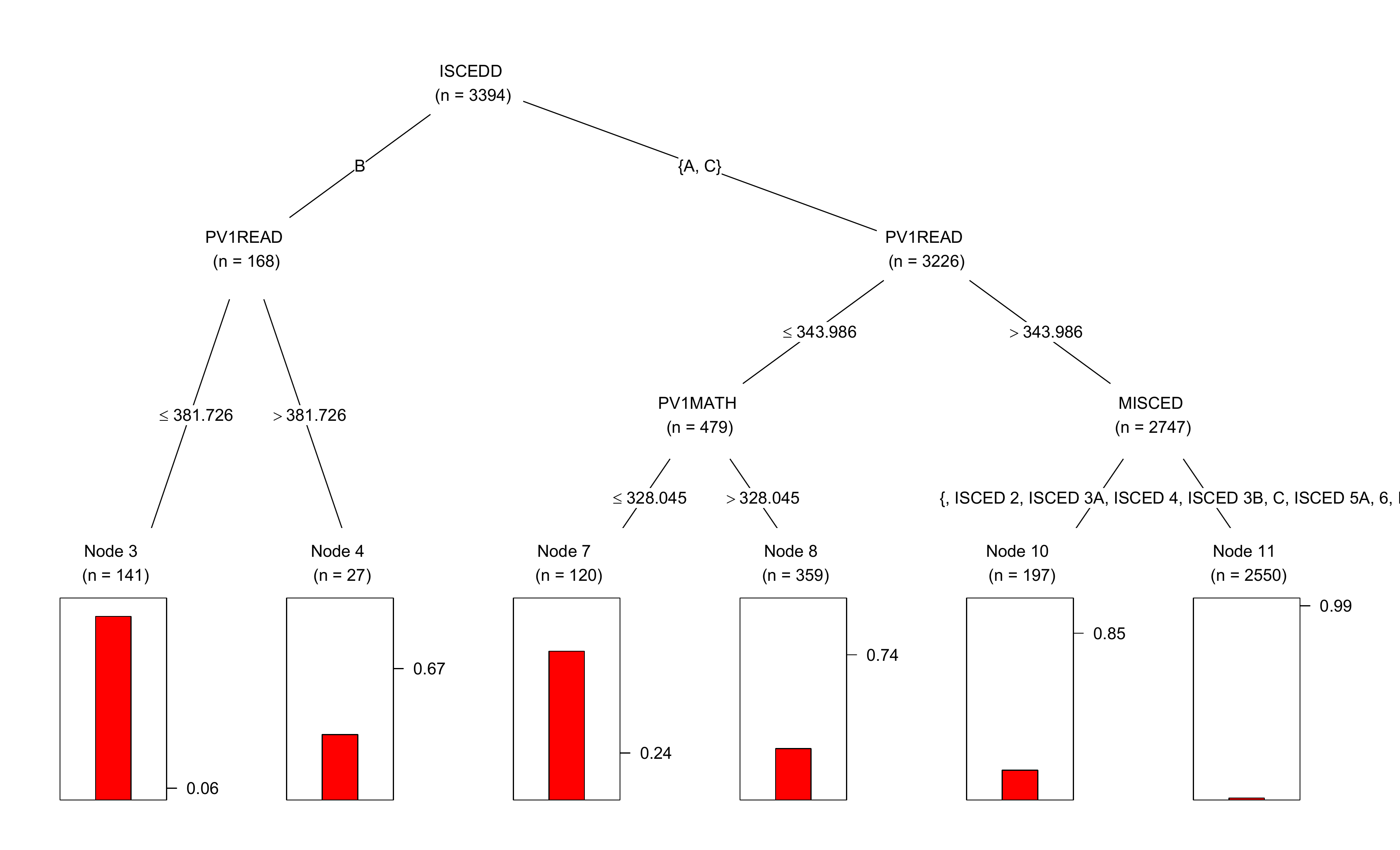}
    \end{subfigure}
    \begin{subfigure}[b]{0.48\textwidth}
    \includegraphics[width=\textwidth]{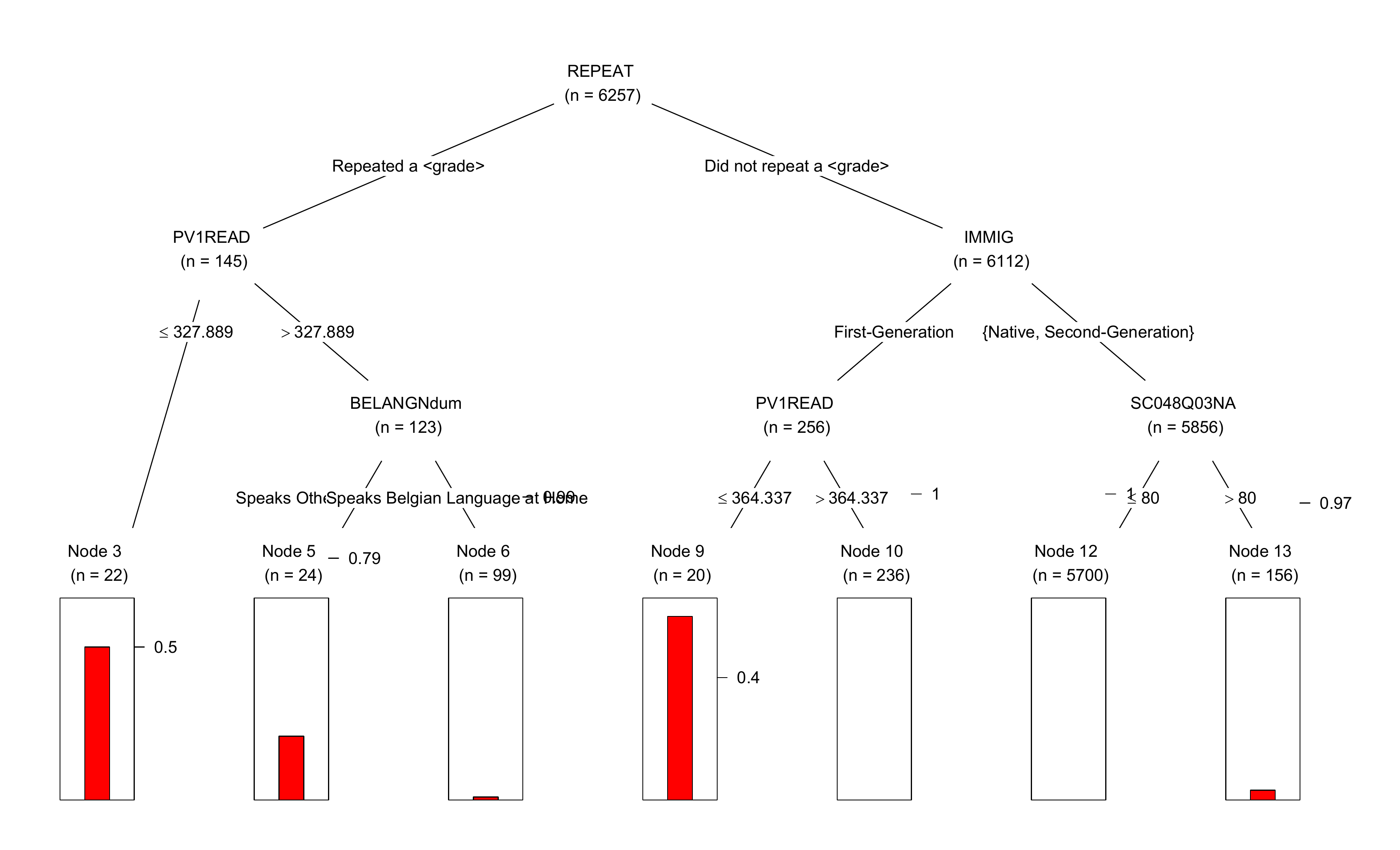}
    \end{subfigure}
    \caption{{\footnotesize(Left) conditional tree for students in grades 7 to 9. (Right) The corresponding tree for students in grades 10 to 12.}}
    \label{fig:ctree_grades}
\end{figure}

\begin{figure}[H]
    \centering
    \begin{subfigure}[b]{0.48\textwidth}
    \includegraphics[width=\textwidth]{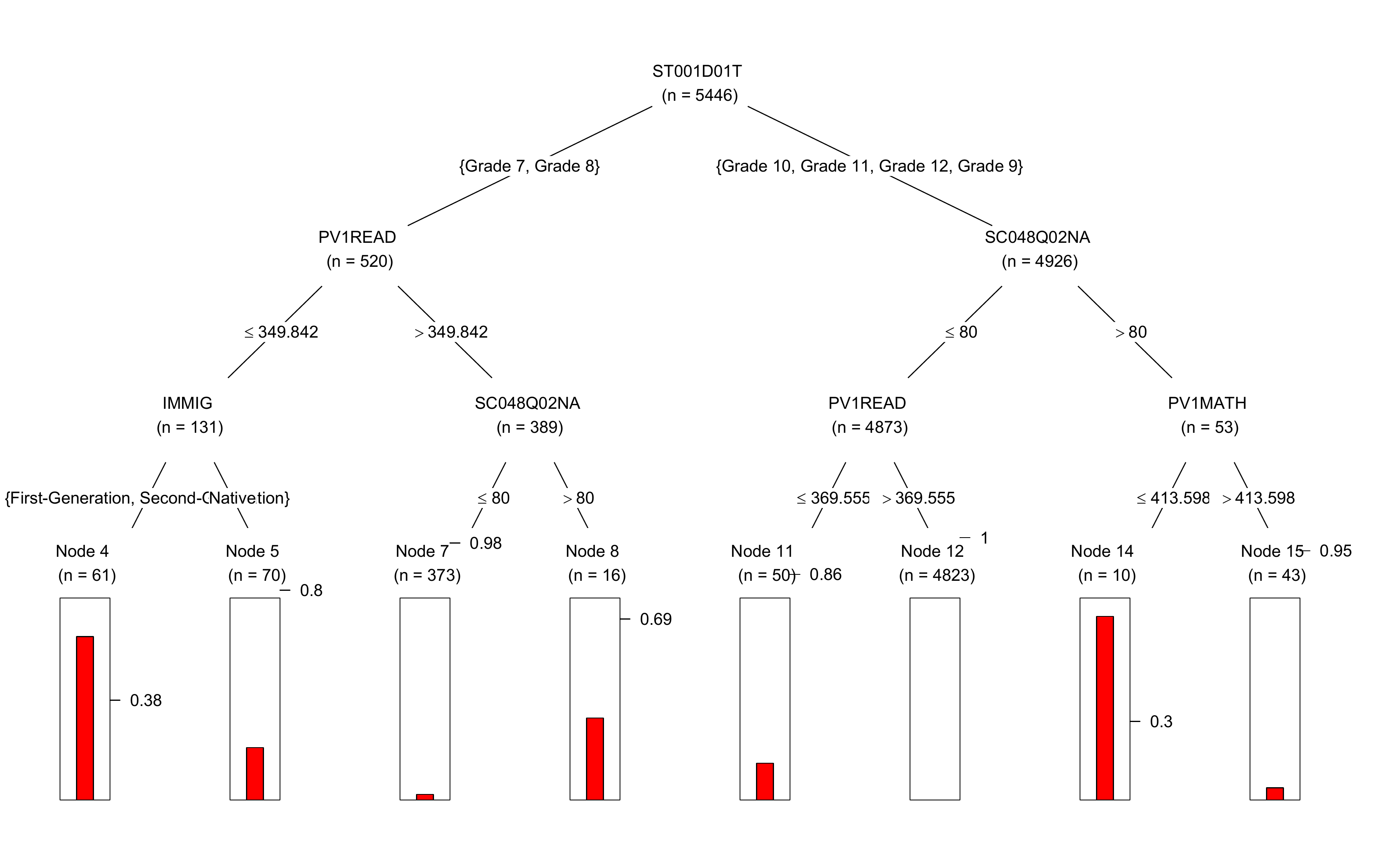}
    \end{subfigure}
    \begin{subfigure}[b]{0.48\textwidth}
    \includegraphics[width=\textwidth]{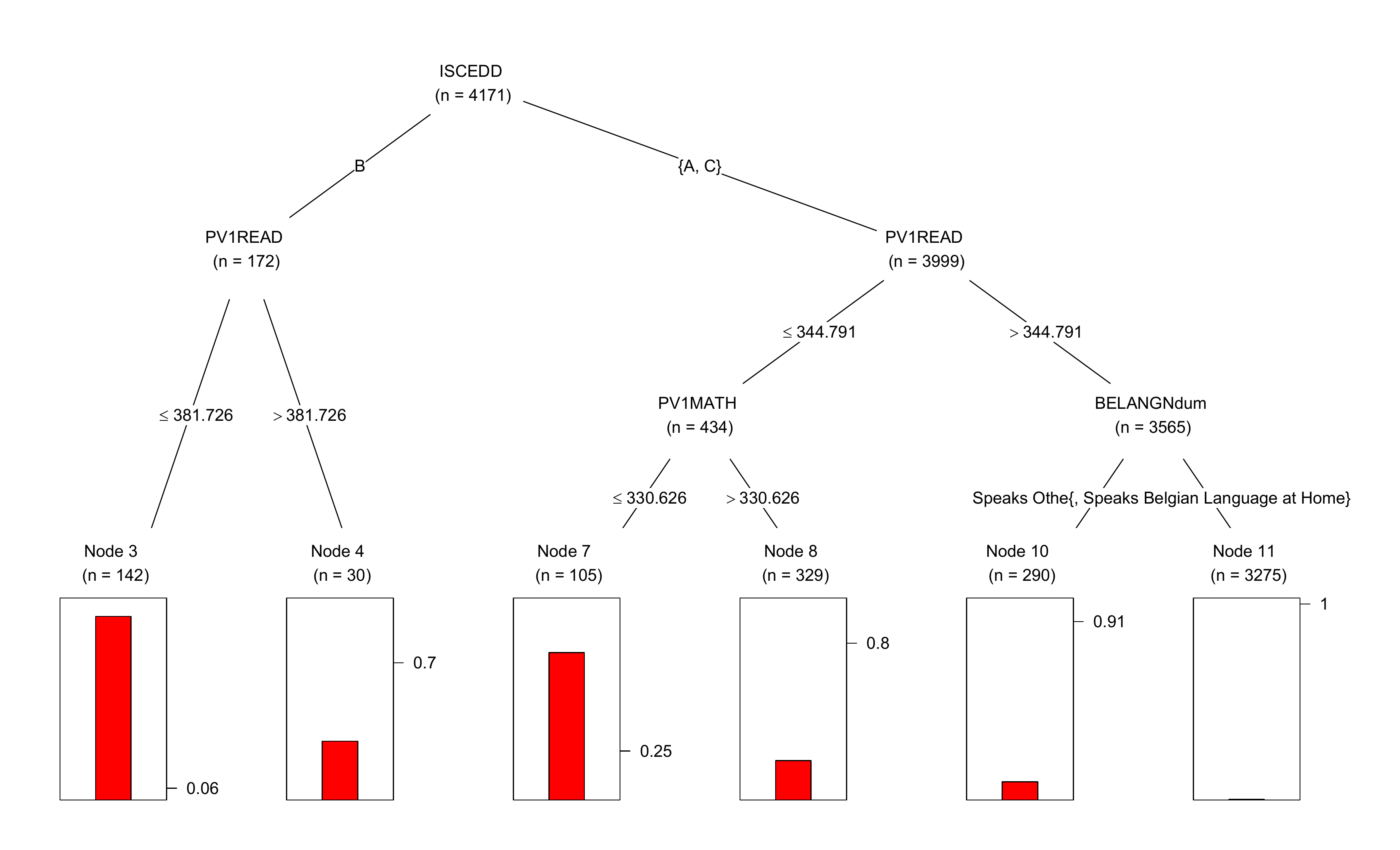}
    \end{subfigure}
    \caption{{\footnotesize(Left) conditional tree for general education. (Right) The corresponding tree for vocational education.}}
    \label{fig:ctree_vocational}
\end{figure}

\end{document}

%% file: Tables/performance.tex
\begin{table}[H]\centering
\small
\caption{Summary Statistics of the Performance of the ML techniques}
\label{table:performance}
\begin{threeparttable}
\begin{tabular}{l*{1}{ccc}}
\toprule
                &     \textit{RMSE} & \textit{MAE} &  $R^2$ \\
\midrule
BART &   58.1799 &   45.7505 &    0.7306 \\
Random Forest &   58.8402 &   46.1451 &    0.7251 \\
\bottomrule
\end{tabular}
\end{threeparttable}
\end{table}

%% file: Tables/stats_outcome.tex
\begin{table}[H]\centering
\small
\caption{Summary Statistics of the Predicted FLS}
\label{table:fls}
\begin{threeparttable}
\begin{tabular}{l*{1}{cccccc}}
\toprule
                &     Mean&       SD&  Minimum&   Median&  Maximum&      N\\
\midrule
Predicted FLS Flanders &   541.4 &   96.8 &    157.3 &   559.9 &   750.0 &     5675\\
Predicted FLS Wallonia &   516.6  &   95.7 &   188.3 &   530.3 &   731.7 &     3976\\
\bottomrule
\end{tabular}
\end{threeparttable}
\end{table}

%% file: Tables/summarystats.tex
\begin{table}[H]
\tiny
\caption{Summary Statistics of the Predictors}
\label{tablesummarystats}
\begin{threeparttable}
\begin{tabular}{l*{3}{{c}{c}{c}}}
\toprule \addlinespace
                & Flanders&         &         & Wallonia&         &         &Difference in Means\\ \addlinespace
                &     Mean&       SD&        N&     Mean&       SD&        N&  p-value\\
\midrule
\               &         &         &         &         &         &         &         \\
\addlinespace
\textit{Student Characteristics}&         &         &         &         &         &         &         \\
\addlinespace
International Grade&     9.72&     0.52&     5675&     9.44&     0.75&     3857&    0.000\\
\addlinespace
Gender          &     1.51&     0.50&     5675&     1.51&     0.50&     3976&    0.976\\
\addlinespace
Age             &    15.85&     0.29&     5675&    15.84&     0.29&     3976&    0.725\\
\addlinespace
Study Track: ISCED Designation &     1.43&     0.81&     5675&     1.24&     0.64&     3976&    0.000\\
\addlinespace
Study Track: ISCED Orientation &     2.09&     1.00&     5675&     1.55&     0.89&     3976&    0.000\\
\addlinespace
Speaks Belgian Language at Home&     0.91&     0.29&     5595&     0.87&     0.34&     3929&    0.000\\
\addlinespace
\               &         &         &         &         &         &         &         \\
\textit{Socioeconomic Status}&         &         &         &         &         &         &         \\
\addlinespace
Educational Resources at Home&     0.27&     0.91&     5585&    -0.15&     0.90&     3930&    0.000\\
\addlinespace
Family Wealth Index (Economic Possessions) &     0.27&     0.74&     5606&    -0.04&     0.85&     3936&    0.000\\
\addlinespace
Number of Books at Home&     3.02&     1.51&     5557&     3.32&     1.53&     3905&    0.000\\
\addlinespace
Immigration Status&     1.20&     0.54&     5512&     1.32&     0.66&     3851&    0.000\\
\addlinespace
Mother's Education (ISCED)&     4.62&     1.39&     5376&     4.63&     1.53&     3774&    0.882\\
\addlinespace
Father's Education (ISCED)&     4.53&     1.44&     5221&     4.52&     1.58&     3654&    0.901\\
\addlinespace
Mother's Job (ISEI)&    47.05&    22.75&     4763&    47.30&    22.10&     3159&    0.635\\
\addlinespace
Father's Job (ISEI)&    46.21&    21.38&     4713&    46.21&    22.16&     3318&    0.996\\
\addlinespace
Parents Emotional Support&    -0.01&     0.96&     5458&     0.01&     0.96&     3801&    0.302\\
\addlinespace
\               &         &         &         &         &         &         &         \\
\textit{Achievement and Attitude}&         &         &         &         &         &         &         \\
\addlinespace
Plausible Value 1 in Mathematics&   521.61&    97.89&     5675&   494.78&    90.71&     3976&    0.000\\
\addlinespace
Plausible Value 1 in Reading&   510.95&   100.67&     5675&   490.46&    95.81&     3976&    0.000\\
\addlinespace
Grade Repetition&     0.24&     0.43&     5457&     0.42&     0.49&     3804&    0.000\\
\addlinespace
Out-of-School Study Time per Week&    14.24&    10.21&     4245&    16.27&    11.61&     3216&    0.000\\
\addlinespace
Mathematics Learning Time at School&   191.28&    86.43&     5273&   219.74&    81.86&     3668&    0.000\\
\addlinespace
Language Learning Time at School&   187.24&    84.63&     5275&   227.10&    83.49&     3665&    0.000\\
\addlinespace
Personality: Test Anxiety&    -0.30&     0.97&     5416&    -0.01&     1.01&     3775&    0.000\\
\addlinespace
Achievement Motivation&    -0.64&     0.83&     5417&    -0.28&     0.87&     3767&    0.000\\
\addlinespace
\               &         &         &         &         &         &         &         \\
\textit{School Characteristics}&         &         &         &         &         &         &         \\
\addlinespace
School Community (Location)&     2.87&     0.81&     5556&     3.24&     1.11&     3681&    0.000\\
\addlinespace
Share of Students With a Different Heritage Language&    16.41&    23.39&     5406&    26.19&    30.29&     2812&    0.000\\
\addlinespace
Share of Students With Special Needs&    19.58&    20.14&     5154&    18.97&    21.94&     3005&    0.208\\
\addlinespace
Share of Socioeconomically Disadvantaged Students &    20.02&    22.11&     5248&    33.06&    30.38&     3120&    0.000\\
\addlinespace
School Size     &   695.86&   331.42&     5405&   771.27&   327.30&     3413&    0.000\\
\addlinespace
Class Size      &    18.89&     5.14&     5592&    21.01&     3.77&     3578&    0.000\\
\addlinespace
Number of Available Computers per Student&     1.25&     0.89&     5270&     0.47&     0.33&     3251&    0.000\\
\addlinespace
Teacher Professional Development&     0.10&     0.90&     5150&     0.11&     1.07&     3433&    0.685\\
\addlinespace
School Autonomy &     0.77&     0.18&     5675&     0.59&     0.21&     3626&    0.000\\
\addlinespace
Shortage of Educational Material&     0.02&     0.87&     5488&     0.21&     0.87&     3429&    0.000\\
\addlinespace
Student-Teacher Ratio&     9.09&     3.21&     5325&     9.14&     2.66&     2847&    0.454\\
\bottomrule
\end{tabular}
\begin{tablenotes} 
\item \footnotesize{\textit{Note:} Summary statistics of all predictors used in the analysis from the PISA 2015 data for Flanders and Wallonia. SD stands for the standard deviation of the variable. The last column shows the p-value of a two-sample t-test for the equality of means across the regions of Flanders and Wallonia. Language has been grouped for Belgium (1=Dutch, 2=French, 3=German, 4=Other).}
\end{tablenotes}
\end{threeparttable}
\end{table}

%% file: Tables/outcome.tex
\begin{table}[H]\centering
\small
\caption{Summary Statistics of the Outcome Variable}
\label{tableoutcome}
\begin{threeparttable}
\begin{tabular}{l*{1}{cccccc}}
\toprule
                &     Mean&       SD&  Minimum&   Median&  Maximum&      N\\
\midrule
Financial Literacy &   541.43&   112.16&    51.81&   555.14&   901.64&     5675\\
\bottomrule
\end{tabular}
\begin{tablenotes} 
\item \footnotesize{\textit{Note:} Summary statistics of Plausible Value 1 in Financial Literacy from PISA 2015 for the Flemish region in Belgium. The OECD constructs ten plausible values for financial literacy using item response theory and latent regression \citep{OECD2017b}. The analysis in this paper is based on Plausible Value 1. SD stands for the standard deviation of the variable.}
\end{tablenotes}
\end{threeparttable}
\end{table}